\documentclass[a4paper,10pt]{scrartcl}
\usepackage[OT1]{fontenc}
\usepackage{amsthm,amsmath,amssymb}
\usepackage{natbib}
\usepackage[colorlinks,citecolor=blue,urlcolor=blue]{hyperref}
\usepackage{graphicx}
\usepackage{mathrsfs}
\usepackage{exscale}
\usepackage{a4wide}
\usepackage{enumerate}
\usepackage{subfigure}
\usepackage{pgfplots}
\usepackage{tikz}
\usepackage{nameref}
\usetikzlibrary{calc}
\usepackage[final]{listofsymbols}
\pgfplotsset{compat=newest}
\newlength\fheight \newlength\fwidth

\usepackage{booktabs}
\usepackage{array}
\newcommand{\bx}{\mathbf{x}}

\newcommand{\bi}{\mathbf{i}}
\newcommand{\bj}{\mathbf{j}}
\newcommand{\bh}{\mathbf{h}}
\newcommand{\g} {g}
\newcommand{\A}{A} 
\newcommand{\gt} {g_t}
\newcommand{\At}{A_t} 
\newcommand{\R}{\mathbb{R}}
\newcommand{\bD}{\mathbf{D}}
\newcommand{\bs}{\mathbf{s}}

\newcommand{\by}{\mathbf{y}}

\newcommand{\trans}{\mathbf{T}}

\newcommand{\transclt}{\Psi}
\newcommand{\transcltD}{\transclt(\mathcal{D})}
\newcommand{\transcltDhat}{\transclt(\widehat{\mathcal{D}})}

\newcommand{\psf}{h}

\numberwithin{equation}{section}
\theoremstyle{plain}
\newtheorem{thm}{Theorem}[section]

\newtheorem{rem}[thm]{Remark}
\newtheorem{ass}[thm]{Assumption}

\newtheorem{lem}[thm]{Lemma}

\opensymdef
\newsym[Anzahl Detektoren]{md}{m_{\text{d}}} 
\newsym[Anzahl Q's]{mQ}{m_{\text{Q}}} 
\newsym[Anzahl der $s_k$, die bestimmt werden]{ms}{m_{\text{d}}} 
\newsym[Anzahl $S_k$]{mS}{m_{\text{S}}} 
\newsym[Anzahl s's, die  zum z\"ahlen verwendet werden]{mc}{m_{\text{c}}} 
\newsym[Anzahl ps f\"ur die Approximation der Qs]{mP}{m_{\text{p}}} 
\newsym[Anzahl D's]{mD}{m_{\text{D}}}
\newsym[]{bright}{p}
\newsym[]{brightx}{\mathfrak{p}}
\newsym[Anzahl Pixel pro Zeile]{m}{n}
\closesymdef

\begin{document}
	
	\begin{center}
		\begin{minipage}{.8\textwidth}
			\centering 
			\LARGE Towards quantitative super-resolution microscopy:\\ Molecular maps with statistical guarantees\\[0.5cm]
			
			\normalsize
			\textsc{Katharina Proksch}\footnotemark[1]\\[0.1cm]
			\verb+k.proksch@utwente.nl+\\
			Faculty of Electrical Engineering, Mathematics and Computer Science, Universiteit Twente, Enschede, The Netherlands\\[0.3cm]
			
			\textsc{Frank Werner}\footnotemark[1]\\[0.1cm]
			\verb+frank.werner@mathematik.uni-wuerzburg.de+\\
			Institute of Mathematics, University of W\"urzburg, Germany\\[0.3cm]
			
			\textsc{Jan Keller--Findeisen}\footnotemark[1]\\[0.1cm]
			\verb+jan.keller@mpinat.mpg.de+\\
		    Max-Planck-Institut für multidisziplinäre Naturwissenschaften, G\"ottingen, Germany\\[0.3cm]
			
			\textsc{Haisen Ta}\\[0.1cm]
			\verb+hta@pysnet.uni-hamburg.de+\\
			Center for Hybrid Nanostructures, Universität Hamburg, Germany\\[0.3cm]
			
			\textsc{Axel Munk}\\[0.1cm]
			\verb+munk@math.uni-goettingen.de+\\
			Institute for Mathematical Stochastics, University of G\"ottingen\\
			and\\
			Felix Bernstein Institute for Mathematical Statistics in the Bioscience, University of G\"ottingen\\
			and\\
			Max-Planck-Institut für multidisziplinäre Naturwissenschaften, G\"ottingen, Germany\\[0.3cm]

		\end{minipage}
	\end{center}
	
	\footnotetext[1]{These authors contributed equally}
	
	\begin{abstract}
Quantifying the number of molecules from fluorescence microscopy measurements is an important topic in cell biology and medical research. In this work, we present a consecutive algorithm for super-resolution (STED) scanning microscopy that provides molecule counts in automatically generated image segments and offers statistical guarantees in form of asymptotic confidence intervals.
To this end, we first apply a multiscale scanning procedure on STED microscopy measurements of the sample to obtain a system of significant regions, each of which contains at least one molecule with prescribed uniform probability. This system of regions will typically be highly redundant and consists of rectangular building blocks. To choose an informative but non-redundant subset of more naturally shaped regions, we hybridize our system with the result of a generic segmentation algorithm. The diameter of the segments can be of the order of the resolution of the microscope.
Using multiple photon coincidence measurements of the same sample in confocal mode, we are then able to estimate the brightness and number of the molecules and give uniform confidence intervals on the molecule counts for each previously constructed segment. In other words, we establish a so-called \textit{molecular map} with uniform error control. The performance of the algorithm is investigated on simulated and real data.
	\end{abstract}

	
	\textit{AMS classification numbers:} 60K35.  \\[0.3cm]

	\date{\today}

	\section{Introduction}
\subsection{Super-resolution microscopy}
In fluorescence microscopy, structures of interest inside a specimen are labeled with fluorescent markers and then imaged using visible light illumination. Only the fluorescence itself and thus the labeled structures are detected, making it possible, for example, to investigate details inside living cells with unrivaled contrast. The tremendous development of super-resolution fluorescence microscopy in recent decades has extended spatial resolution beyond the diffraction limit of conventional microscopy to the nanometer scale.

All super-resolution light microscopy concepts rely on distinguishing fluorophores locally by consecutively transferring them between a dark (non-fluorescent) and a bright (fluorescent) state using light to induce these transitions (\citet{hell07, sahl17}). The transitions between these states can be performed either in a spatially controlled or in a stochastic manner, with the latter denoted here as single-molecule switching (SMS) microscopy (\citet{betzig2006}). In both approaches only a small subset of molecules is left in the bright state at each measurement step and the final image is assembled by repeating the experiment many times. A well established spatially controlled method uses stimulated emission depletion (STED) (\citet{hell94, klar00}). Thereby a red-shifted light spot featuring a central intensity minimum is co-aligned with the excitation light spot. It induces strongly saturated stimulated emission, effectively inhibiting the fluorophores from emitting fluorescence in the periphery of a focused excitation light spot. The very small spot of effectively allowed fluorescence emission can be scanned over the sample, where each scanning position in a rectangular grid corresponds to a pixel in the final image. For example, the STED principle has been used in the past to reveal 
the distribution of synaptic proteins in living mice (\citet{masch18})
or the dynamics of membrane lipids in living cells (\citet{eggeling09}). Recently, a combination of stochastic switching and excitation light patterns with at least one isolated intensity zero, called MINFLUX, was used to achieve isotropic resolution on the order of a few nanometers (\citet{balzarotti16}).

From a statistical perspective, the recovery of spatial intensity and specimen distribution from super-resolution fluorescence microscopy images leads to sophisticated convolution models with Poisson or Binomial data distributions, which in themselves present a number of challenges (see, e.g., \citet{aem15, hw16, msw20, kulaitis} and references therein).

In many biological contexts, however, it is not only the precise localization of structures that is of interest, but also the determination of the exact number of fluorescent markers at a given location, especially if this number can be related to the local number of proteins or other biological targets of interest. Such quantitative knowledge of target structures at the nanoscale has the potential to greatly improve the understanding of many biological processes. Knowledge of the absolute number of molecules can provide the basis for structural models of protein complexes or determine thresholds for the number of molecules required to produce a particular effect. For example, estimating the number of constituent proteins in kinetochores reported unexpectedly high numbers of proteins present (\citet{coffman11}), whereas quantifying the number of proteins used for flagellar regeneration helped refine models for flagellar assembly (\citet{engel09}).

\subsection{Towards molecule counting}
In general, the mean recorded fluorescence signal in a microscope is proportional to the number of active fluorescent markers. If we denote by $f_N$ a spatial function assigning each location (or pixel) the corresponding number of markers, and by $f_p$ their corresponding brightness (i.e., probability to emit a photon after an excitation pulse was applied), then the observed quantity is mathematically given by a spatial convolution of the product $f_N \cdot f_p$ with the so-called point spread function (PSF) $\psf$, which is determined by the microscope (see Section~\ref{sec:model} below for details). This already shows that the observations cannot be readily used to infer the absolute number of markers, since the brightness of a fluorophore depends on the local environment and thus $f_p$ is unknown and not constant. If the microscope could guarantee that each fluorophore was perfectly separated, counting fluorophores would be a trivial task, but in almost all current applications this is not the case. Robust statistical modeling of the contributions of the fluorophores to the acquired image data, and, in particular, a method for calibrating the local brightness of molecules, are therefore required. This calibration is ideally performed during the measurement itself, since the molecular brightness also depends on the microscope configuration used and the sample conditions.

In the super-resolution techniques that leave only isolated markers in the bright state (as it is, e.g.\ the case in SMS microscopy), counting can be performed by a careful analysis of the localization events (see e.g. \citet{Lee17436, RollinsE110, Hummeretal2016, staudtetal2020}). From a statistical perspective this is mainly based on the \textit{temporal} Markovian dynamics of the transition between bright and dark states of the fluorophores (\citet{PatelAOAS2019, GabittoAOAS}) and requires detailed knowledge about the fluorescent state transition kinetics.

In case of \textit{spatially} separating super-resolution microscopy, e.g.\ for STED as considered in this work, a similar approach is not possible. Instead, the single molecule brightness has traditionally been estimated by observing single photon bleaching steps, which is challenging in dense samples. For the case of repeatedly activatable markers one can also determine an excess variance, which was used to determine the molecule brightness (\citet{Parketal2005, Frahmetal2019}). Despite its practical relevance, stochastically sound methods for counting molecules from scanning microscopy images remain elusive until nowadays.

\subsection{Our approach}
\begin{figure}[!b]
	\centering
	\includegraphics[width=0.9\textwidth]{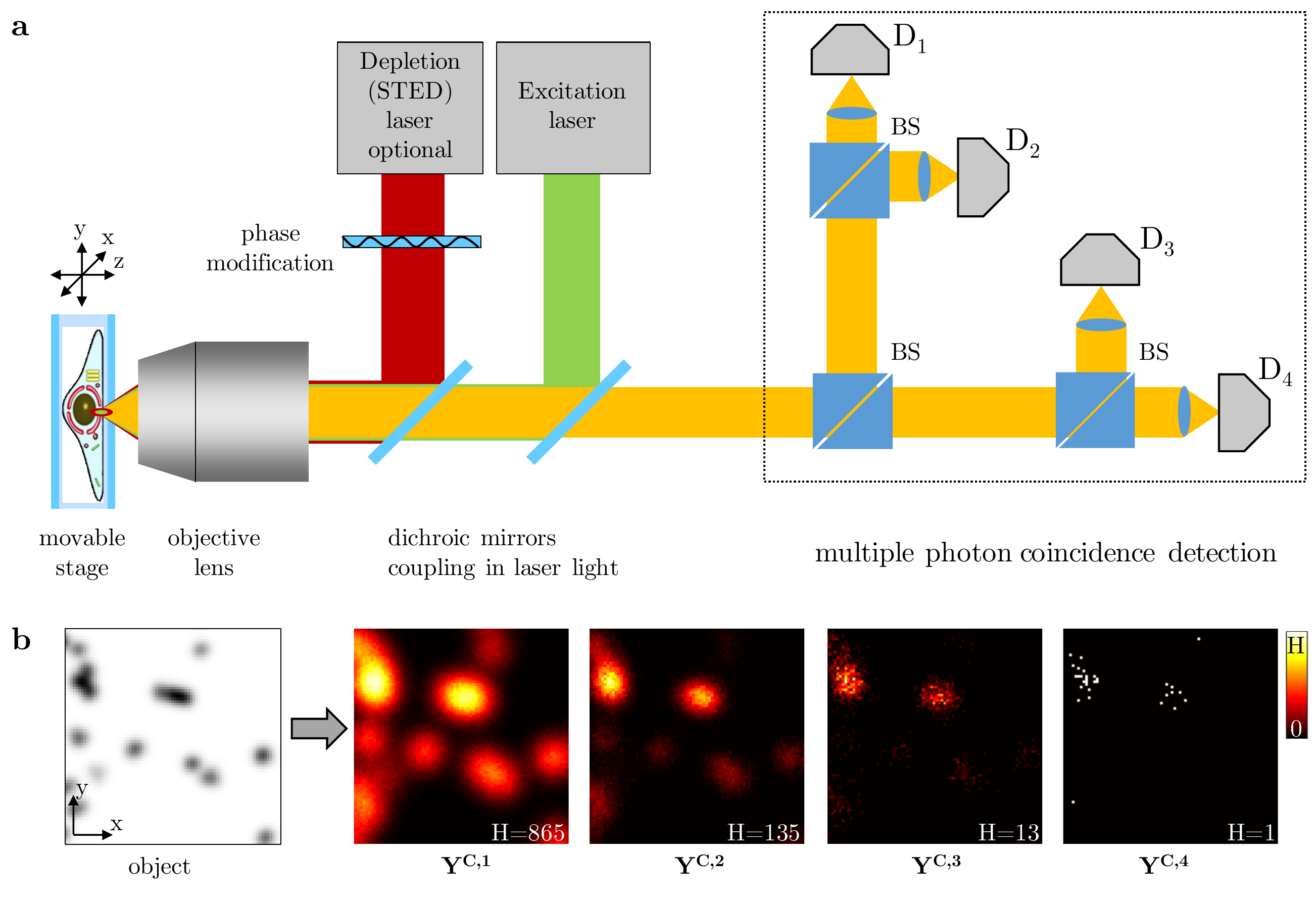}
	\caption{\footnotesize\textbf{Microscope scheme and example images for detecting multiple photon coincidences.} \textbf{a} Schematic of a confocal/STED fluorescence light microscope capable of detecting multiple photon coincidences. Excitation, depletion and fluorescence light are combined with dichroic mirrors. A specific phase distribution is imprinted on the depletion beam in order to create a doughnut shaped intensity distribution of depletion light at the focal point. Effectively, the spot in the sample that is allowed to fluoresce is reduced to a sub-diffraction extent. In the detection unit, the fluorescence light is split by three beam splitters (BS) in four equal intensity paths and directed on four identical detectors ($D_k, k=1,..,4$). Such a microscope has been built and used for antibunching microscopy in (\citet{t15}). \textbf{b} Example object (molecule density) and corresponding simulated photon coincidence measurement images $\mathbf Y^{\text{C},k}, k=1,..,4$ in confocal mode (excitation pulses per pixel $t=3000$), see Section~\ref{sec:model} for details on model and data. The scaling of the colorbar is indicated by the value H in each image.}
	\label{fig:experimental_setup}
\end{figure}

In this work, we will address this issue by means of a different path. To this end, we will perform counting based on the photon emission statistics, specifically the number of photons that are emitted simultaneously. This has recently emerged as a tool for intrinsically calibrating the molecular brightness and to infer on the number of molecules present. The physical effect affecting the photon emission statistics is called photon antibunching, i.e., an excited fluorophore cannot emit more than one photon during the lifetime of the excited state. First measurements showed that the number of markers can be inferred from the photon emission statistics (\citet{t10}). Termed antibunching microscopy, it has been implemented later in STED mode and could for example count the number of internalized receptors within small vesicles in HEK293 cells (\citet{t15}). This technique is able to account for locally varying molecular brightness: Given that the event of observing a single photon has a probability proportional to $f_N {\cdot} f_p$ as argued above, the probability that two photons are measured simultaneously must be proportional to $\left(f_N \cdot f_p\right)^2$ minus the physically impossible contribution that both photons stem from the same molecule, which is itself proportional to $f_N {\cdot} f_p^2$. So, after a linear transformation, we obtain observations proportional to $f_N {\cdot} f_p$, $f_N {\cdot} f_p^2$, ..., which in principle allows decoupling the (local) number of markers $f_N$ from their brightness $f_p$. For details, see Theorem~\ref{Le:trans} below and its proof in the supplement. A scheme of an antibunching STED microscope is shown in Figure~\ref{fig:experimental_setup}. It provides simultaneous detections of multiple photons, i.e., the measurement of photon coincidences. State of the art single photon detectors (avalanche photo diodes) still feature considerable times where they are insensitive after each photon detection (on the order of ~50 nanoseconds), rendering them incapable of detecting photon coincidences. This limitation is effectively overcome by parallel detection, i.e., by splitting the emitted light equally among multiple detectors.
	
At each scan position (the scan covers the whole sample equally) a fixed number $t$ of excitation light pulses is applied and the statistics of detected number of photons are recorded. This can be seen as repetitions of a multinomial experiment with an additional twist that multiple photons arriving at the same detector after each excitation light pulse are only registered as a single detection event. The accessible observable is indeed only the number of active detectors for each excitation light pulse, which depends on the number of emitted photons and the number of the available detectors. 
In Section~\ref{sec:model}, we develop an explicit multinomial model that encodes the photon emission and detection probabilities and is given by convolutions and products of the local brightness, the local number of markers, and the effective point spread function of the microscope itself. The accuracy  of the estimation of the multinomial probabilities of our model increases with increasing number $t$ of excitation light pulses. Therefore, our theoretical considerations assume an asymptotic viewpoint as this number tends to infinity. Typical numbers of $t$ in applications easily reach $t=10,000$, such that the approximation by the asymptotic results is rather accurate.   
Utilizing this model, measuring multiple photon coincidences allows for inference on the (local) number of markers, as the simultaneous arrival of two photons implies that the two photons must originate from at least two different fluorophores. We will show later on, that the empirical distribution of the number of active detectors depends uniquely on the number and location of molecules present. An estimation of the local molecular number and brightness by a penalized maximum likelihood based reconstruction algorithm (without statistical guarantees) was implemented in \citet{t15}. Besides lack of statistical guarantees for the local molecular numbers, the statistical model therein is limited to two simultaneously arriving photons at most.

\subsection{Our contribution}
As the goal of counting the number of fluorescent markers in a sample would greatly benefit from both being able to calculate error bounds on the numbers and considering higher order photon coincidences,
in this work, we will extend the existing approach in these directions. First of all, we derive a detailed and sound statistical model for the observations obtained from the antibunching microscope including contributions of arbitrarily high photon coincidences. One major contribution is that instead of globally estimating the number of markers inside the specimen, we construct a so-called \textit{molecular map} with uniform error control. This is a collection of family-wise error rate (FWER) controlled, distinct segments in the specimen together with uniform confidence intervals for the number of markers contained in each segment. The diameter of these segments can be of the order of the resolution of the microscope. To derive this molecular map with a given total error level $\alpha \in \left(0,1\right)$, the method performs three steps:

\begin{description}
	\item[S1:] \textbf{Segmentation}. We first construct a segmentation of the image space such that each segment contains at least one molecule with uniform probability at least $1-\alpha/2$ and, at the same time, has a suitable shape. All image pixels that are not contained in any of the segments are no longer considered in steps S2 and S3.
	\item[S2:] \textbf{Estimation.} Given $M$ (say) segments from S1, we estimate (locally) the number of molecules, i.e. segment by segment.
	\item[S3:] \textbf{Confidence}. Based on a central limit theorem, we construct $1-\alpha/(2M)$ confidence intervals (CIs) for each of the $M$ local (segment-wise) numbers of molecules.
\end{description}

The outcome of this procedure is a collection of distinct segments, estimated marker counts in these segments and confidence intervals such that asymptotically (as the number $t$ of excitation light pulses tends to infinity) the following statement is true (see Theorem~\ref{thm:uglythm}):
\begin{align}\label{eq:coverage}
	\mathbb P \left[\text{Each CI contains the correct number of markers in its segment}\right] \geq 1-\alpha.
\end{align}
The provided MATLAB\textsuperscript{\textregistered} code implements the approach described above, and we investigate its performance in numerical simulations as well as real data examples in Sections~\ref{sec:numerical} and \ref{sec:origami}.

Let us discuss some immediate issues and the rationale behind this three step approach. In principle, the measured photon coincidences do allow for a pixel-wise estimation of the number of markers, and hence also for pixel-wise confidence statements. This is also possible with the above three step approach by considering each pixel as a separate segment. However, to obtain a uniform coverage as in \eqref{eq:coverage}, the corresponding confidence intervals (e.g.\ as constructed in step S3) will then be unfavorably large due to the large number of pixels (in practical applications on the order of $10^5-10^6$) and its corresponding multiplicity correction. 
To overcome this burden, we exploit the fact that in many samples the molecules are concentrated in parts of the measurement volume and our algorithm estimates on relevant regions only. Consequently, it neglects a substantial fraction of all pixels before performing the estimation. If the segments are still chosen reasonably small (e.g.\ in the order of the microscope resolution rather then single pixels), local information is maintained while at the same time the number of confidence statements to be made is reduced significantly. Therefore, a reasonable segmentation in step S1 will strongly ease the estimation and confidence procedure afterwards. Such a segmentation will be achieved by a hybridization of an established segmentation approach from image processing such as, e.g., the watershed segmentation, with a hot spot detection procedure that provides rigorous statistical guarantees such as the \textbf{M}ultiscale \textbf{I}nverse \textbf{SCA}nning \textbf{T}est (MISCAT) procedure introduced in \citet{pwm16}, which is a method based on multiple statistical hypothesis testing. Such a hybrid approach to segmentation is new and offers a lot of flexibility in the choice of methods while providing a strong theoretical justification. Note, that even though the number $M$ of segments in step S2 is random, this does not cause problems in step S3, as it can be considered deterministic conditional on the data used for step S1. 
Moreover, in our asymptotic considerations, the number of pixels, which is an upper bound for $M$, is fixed. 

\begin{figure}[!b]
	\centering
	\includegraphics[width=\textwidth]{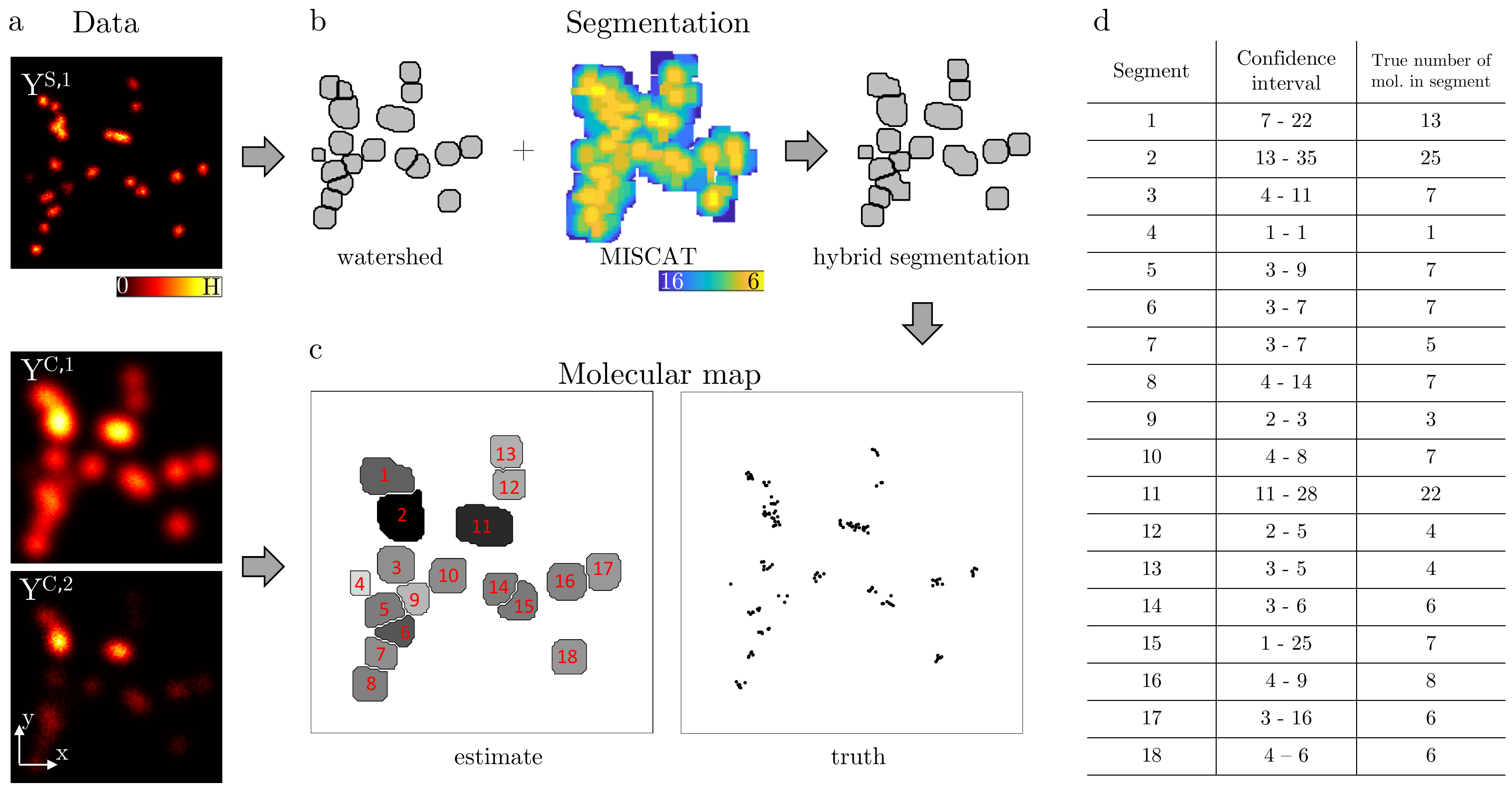}
	\caption{\footnotesize \textbf{Three step analysis workflow.}  \textbf{a} Simulated one photon STED image with high spatial resolution (top left) and one and two photon detection images in confocal mode (center left and bottom left) for the example object depicted in Fig.~\ref{fig:experimental_setup}b. \textbf{b} The STED image $Y_1^S$ is used for a combined watershed and MISCAT segmentation that results in an effective hybrid segmentation, where each segment contains at least one molecule with uniform probability. Color-scale for MISCAT is inverse to box area and smaller boxes are drawn on top of larger boxes. \textbf{c} Using the one and two photon statistics obtained in confocal mode $\mathbf Y^{\text{C},1}$, $\mathbf Y^{\text{C},2}$ and the hybrid segmentation in b, the number of molecules and confidence bounds can be obtained for each segment. The segments are enumerated (red) and the fill color (on a gray scale) indicates the estimated local molecule density within a segment. The true location of markers is shown on the right. \textbf{d} Segment number, confidence bounds ($\alpha=0.1$) and true number of molecules for each of the 18 obtained segments.}
	\label{fig:workflow}
\end{figure}

The three step approach is visualized in Figure~\ref{fig:workflow}. In the example shown there, the images have $512\times512= 262,144$ pixels, but $M$, the number of identified segments containing markers, is only 18. 

	We investigate the introduced methodology both in simulations and on experimental data. The former gives a precise description of the abilities and limitations, and the latter shows the high practical value of our approach.

The remainder of the paper is organized as follows: In Section~\ref{sec:model} we state the statistical model for raw photon counts in antibunching microscopy and their relation to the local number of markers inside the sample. The technical derivation of this model is deferred to Section B.1 of the  supplement. Sections~\ref{sec:segmentation} and \ref{sec:estimation} are devoted to the three steps of our inference algorithm. Detailed numerical simulations are provided in Section~\ref{sec:numerical}. In Section~\ref{sec:origami} we discuss the applicability on real world data. We end with a short conclusion in Section~\ref{sec:discussion}.

\section{Modelling, notation and prerequisites}\label{sec:model}

Throughout this work, we denote by $[n]$ the set $\{1,\ldots,n\}$. Vectors and multi-indices will be written in bold face, where, with a slight abuse of notation, we denote the vector of component-wise ratios $i_k/n_k$ by $\bi/\mathbf{n}$.

In the following, we denote by $N$ the  total (unknown) number of fluorescent markers in our specimen of interest. W.l.o.g. we can assume that all markers are  contained in the unit square $\left[0,1\right]^2$, which is discretized by the grid 
$
\bx_{(i_1,i_2)} = \bx_{\bi} = \left(\frac{i_1}{n}, \frac{i_2}{n}\right), \; \bi \in \left[n\right]\times \left[n\right].
$
The $j$th marker has a position $\bx_{\psi(j)}$ on this grid (where the true position is just assigned to the closest grid point), which is encoded in terms of the mapping $\psi:[N]\to[n]\times[n]$ that assigns to the $j$-th marker, $j\in[N]$, its position on the grid. Furthermore, the $j$-th marker has an individual brightness $\bright_j$. The individual brightness of a marker $j$ is its probability to collect a photon after an excitation light pulse focused to $\bx_{\psi(j)}$ was applied, see, e.g., \citet{aem15}.

The antibunching microscope shown in Figure~\ref{fig:experimental_setup} works as follows. The specimen is scanned spatially along the grid $\bx_\bi,\bi\in [n]\times[n]$. For each grid point $\bx_\bi$, the following experiment is repeated a total number of $t$ times: A short excitation pulse with a duration much shorter than the excited state's lifetime is applied to the sample (focused on the current grid point $\bx_\bi$), and afterwards the fluorescence is recorded at the detectors. Between two experiments, there is a certain waiting time, which takes into account the dead-time of the detectors after recording a photon (typically around $100$ nanoseconds) and the typical fluorescence lifetime of the markers (typically $\leq 10$ nanoseconds). In each experiment, the number of detected photons, $k$, is recorded. 

It is important to note that  a pulse centered at the point $\bx$ will also illuminate neighbouring grid points due to diffraction, however, with a lower intensity. This implies that the probability $\brightx_j(\bx)$ to detect an emission from the $j$-th marker in a single excitation pulse when scanning at position $\bx\in[0,1]^2$ is given by 
\begin{align}\label{eq:bright}
	\brightx_j(\bx) =\bright_j\cdot\psf(\bx-\bx_{\psi(j)}).
\end{align}
Here, $\psf$ is point spread function of the microscope, which reaches its maximal value at 0, and therefore a detection at the correct position is most likely. However, relation \eqref{eq:bright} shows that markers collectively contribute to the measured signal at a specific scanning position $\bx$, even though they are located at different grid points. The shape of $\psf$ depends on various experimental parameters in the microscope. As a rough guidance, $\psf$ has a larger extent when using conventional (i.e. not super-resolved) microscopy (as more surrounding markers are also excited), and a smaller extend in case of STED measurements corresponding to a smaller or larger resolution. An often used measure describing the shape of $\psf$ and also of the microscope's resolution is its full width at half maximum (FWHM), cf. \citet{kulaitis} for an explanation in statistical terms. In case of a Gaussian peak PSF $\psf$ with variance $\sigma^2$, one has FWHM $=2\sqrt{2\log 2}\sigma$.

As a consequence of \eqref{eq:bright}, we obtain by superposition a convolution model
\begin{equation}\label{eq:conv}
	g\left(\bx\right) = \sum_{j=1}^N \brightx_j(\bx) = \int_{\left[0,1\right]^2} f_N(\mathbf{y}) f_p\left(\mathbf{y}\right) h\left(\bx-\mathbf{y}\right) \,\mathrm d \mathbf{y},
\end{equation}
where $g\left(\bx\right)$ denotes the probability to detect at least one photon when scanning at $\bx \in \left[0,1\right]^2$, $f_N: [0,1]^d \to \R$ denotes the spatial number of markers and $f_p : [0,1]^2 \to \R$ their brightness at $\bx\in \left[0,1\right]^2$. Note, that formally $f_N$ is a (finite) sum of dirac-measures (located at the molecule positions) rather than a function, and hence the integral in \eqref{eq:conv} is in fact just a sum. The same applies to all integrals involving $f_N$ below. As an alternative, one could model $f_N$ as a density as well, which underlies all classical (continuous) convolution models in microscopy. Using the notation in \eqref{eq:conv}, we stay in-line with these models, and essentially all subsequent derivations work in both models. 


The convolution model in \eqref{eq:conv} is most commonly used for standard fluorescence microscopes to recover the product $f_N\cdot f_p$, see, e.g.\ \citet{aem15}. However, our modified microscope sketched in Figure~\ref{fig:experimental_setup} does not only measure the total number of photons when scanning at $\bx_\bi$, $\bi \in[n]\times[n]$, but furthermore measures the number of coincidences, i.e., $k$ detected photons at the same time for $0 \leq k \leq \md$, where $\md$ denotes the number of parallel detector units present in the microscope (see Figure~\ref{fig:experimental_setup}). This allows to finally decouple $f_N$ and $f_p$. The data for the current grid point $\bx_\bi$ consequently consists of integers $ Y_\bi^{k}$ being the number of $k$-photon events at $\bx_i$, $0\leq k\leq\md$. This immediately implies that our (ideal) data will consist of (pixel-wise) multinomial observations, i.e.
\begin{equation}\label{eq:data}
	\mathbf{Y}_\bi=\left(Y_\bi^0,...,Y_\bi^{\md}\right)\sim \mathcal M \left(t,E_0\left(\bx_\bi\right), ..., E_{\md} \left(\bx_\bi\right)\right), \qquad \bi\in[n]\times[n],
\end{equation}
where $\mathcal M$ denotes the multinomial distribution with a total number of $t$ experiments per image pixel, and the numbers $E_{k}(\bx_\bi)$ denote the probabilities to observe $k$ photons at position $\bi$. Note, that in the notation of \eqref{eq:conv}, we have
	$	g(\bx_{\bi}) = \sum_{k \geq 1} E_k(\bx_{\bi})$, and due to multinomiality it holds $t=\sum_{k=0}^{\md} Y_i^{k}$
	\, for each pixel $\bi \in [n]\times[n]$.
	
	Throughout this work, we assume that the observations for different grid points $\bx_\bi$ are independent, which means that we assume that the observations $\mathbf{Y}_\bi, \bi\in [n]^2$ are independent, multinomially distributed random variables. This is a reasonable assumption as long as markers do not come so close to each other that they interact.
	Note that model \eqref{eq:data} does not incorporate external noise sources such as detector read-out errors. However, such statistical thinning as it might be caused by loss of photons in the detectors can be included by re-defining the brightness $\bright_j$ of the individual molecules, cf. \citet{msw20}. In conclusion, the model \eqref{eq:data} can be considered as highly accurate if the number of repetitions per pixel, $t$, is of the order of several thousands, which corresponds to typical experimental conditions.
	
	A detailed model for the probabilities $E_{k}, k=0,\ldots,\md$, in dependence on the unknown quantities of interest such as the local number of markers $f_N$ and their brightness $f_p$ 
	is developed in the supplementary material. To describe it briefly, let us introduce the function $\bs:[0,1]^2\to\mathbb{R}^{\ms}$, defined by
	\begin{equation}\label{eq:S}
		\bs(\bx_{\bi})=\left(\sum_{j=1}^N(\brightx_j(\bx_{\bi}))^k\right)_{k=1,\ldots,\ms} = \left(\int_{\left[0,1\right]^2} f_N(\bx) f_p(\bx)^k h(\bx-\bx_{\bj})^k \,\mathrm d \bx\right)_{k=1,\ldots,\ms}.   
	\end{equation}
	Then, $\bs$ can be related to the probabilities $\mathbf E = \left(E_k\right)_{k=0,...,\md}$ -- which can themselves be estimated from the available observations -- as stated in the following theorem.
	\begin{thm}[Statistical model for the antibunching microscope]\label{Le:trans}
		Let $\md$ denote the number of detectors, $N$ the total number of fluorescent markers in the specimen and $\kappa=\kappa(\bx)$ the number of emitted photons in one excitation pulse at position $\bx$.
		Assume that $N>\md$.
		Let 
		$
		\mathbf{E}(\bx_{\bi}):=(E_0\left(\bx_\bi\right), ..., E_{\md} \left(\bx_\bi\right))
		$
		denote the multinomial probabilities defined in model \eqref{eq:data}. Assume that, for any $\bx\in[0,1]^2$, there exists a $\gamma=\gamma(\bx)\in(0,1)$ such that for any $\kappa>m_d$ it holds that
		\begin{align}\label{eq:gamma}
			\mathbb{P}(\text{more than }\kappa(\bx)\; \text{photons emitted at position }\bx\; \text{during one pulse})\leq\gamma(\bx)^{\kappa}.
		\end{align}
		Then there exists an explicitly known differentiable, invertible map $\trans:[0,1]^{\md}\to\mathbb{R}^{\ms}$ such that
		\begin{align}\label{eq:modelbias}
			\mathbf{E}(\bx_{\bi}) =\trans( \bs(\bx_{\bi}))+O\left(\gamma^{\md+1}\right)\quad\text{ as}\;\gamma\to0
		\end{align}
		for all $\bi \in [n]\times[n]$.
	\end{thm}
	\noindent For an explicit formula for $\trans$, see the proof of Theorem~\ref{Le:trans}. 
	
	\noindent Note that assumption \eqref{eq:gamma} will be satisfied as soon as the local number of molecules around a position $\bx$ and/or their brightness is not too large compared to the number of detectors $\md$. As a consequence, for sufficiently large $\md$, we can neglect the remainder term in \eqref{eq:modelbias} and obtain (approximately) a non-linear forward model
	\begin{equation}\label{eq:ip}
		\mathbf E = \left(E_k\right)_{0 \leq k \leq \md} = F \left(f_N, f_p\right),
	\end{equation}
	with $F$ given by the concatenation of $\trans$ and the mapping $\left(f_N, f_p\right) \mapsto \bs$ (recall \eqref{eq:S}), from which we aim to determine local information on the spatial number $f_N$ (and the spatial brightness $f_p$) of molecules from empirical measurements $\mathbf Y$ as in \eqref{eq:data}. This can be seen as a nonlinear inverse problem, which is also ill-posed due to convolution with the PSF $\psf$.
	
	Note that the model \eqref{eq:ip} does not take background contributions into account, e.g.\ from out-of-focus planes. This is for simplicity mostly, as the following considerations can immediately be adjusted to a spatially varying background intensity $\lambda$, which, however, leads to even more technical results. In principle this would allow to determine not only $f_N$ and $f_p$ from the available data, but also $\lambda$. In practice, this corresponds to a highly underdetermined problem, which is why typically the background intensity is pre-estimated and then used to correct the data. To keep the theory concise, we have therefore decided to neglect all background contributions in the main document for the sake of simplicity. A more refined model including the estimation of $\lambda$ is presented in the implementation (cf. Section~\ref{sec:origami}).
	
	In practice, we are able to obtain the measurements $\mathbf Y$ in different imaging modes. As shown in Figure~\ref{fig:experimental_setup}, the microscope has an additional STED laser, which can either be turned on or off, yielding different effective PSFs $\psf$. This allows us to image the specimen once via classical confocal microscopy (where the STED laser is turned off) and once via super-resolution STED microscopy.
	In total, this means that we collect and analyze both STED data $\mathbf Y^{\text{S}} = \left(\mathbf Y^{\text{S},0},...,\mathbf Y^{\text{S},\md}\right)$ consisting of matrices containing the pixel wise $k$-photon counts $\mathbf Y^{\text{S},k} = \left(Y_{\bi}^{\text{S},k}\right)_{\bi \in [n]\times[n]}$ as well as confocal data $\mathbf Y^{\text{C}} = \left(\mathbf Y^{\text{C},0},...,\mathbf Y^{\text{C},\md}\right)$ consisting of similar matrices $\mathbf Y^{\text{C},k} = \left(Y^{\text{C},k}_{\bi}\right)_{\bi \in [n]\times[n]}$. The properties, advantages and disadvantages of both imaging modes will be discussed in more detail later on.
	
	\section{Segmentation}
	\label{sec:segmentation}
	
	If we tried to construct pixel-wise confidence intervals for the local number of molecules everywhere in the image (comparable to \eqref{eq:coverage}), we would not obtain meaningful results due to the large number of pixels, as the necessary multiplicity adjustments would inflate the results. Therefore, it is important to first select regions of interest (RoIs) in the image, on which we afterwards analyze the local number of molcules. The aim of this section is to introduce a both intuituve and statistically rigorous data driven segmentation of the given image into 'active' regions (i.e. containing molecules)  and 'inactive' regions (i.e. containing no or only very few molecules). To make this useful, we aim to select a system of RoIs such that
	\begin{description}
		\item[(R1)] all interesting clusters of molecules are contained in one of the RoIs,
		\item[(R2)] with high probability each RoI contains at least one molecule,
		\item[(R3)] the RoIs do not intersect (i.e. form a valid segmentation), 
		\item[(R4)] the RoIs are reasonably small and 
		\item[(R5)] the RoIs have suitable shapes.
	\end{description}
	
	A standard approach to this problem would be the usage of a standard data-driven segmentation algorithm on the STED data $\mathbf{Y}^{\text{S},1}$ (offering a much better resolution compared to a confocal image), which -- using suitable tuning parameters -- hopefully yields reasonable RoIs satisfying (R1) and (R3)--(R5). However, strong statistical guarantees such as (R2) do typically not hold. On the other hand, systems of sets satisfying (R2) are often overlapping and thus (possibly highly) redundant, i.e.\ violate (R3), and furthermore do not satisfy (R5). Our approach is therefore to profit from the strengths of two different approaches, creating a hybrid version that inherits the positive aspects of both ingredients. A theoretical guarantee of property (R1) is very difficult and is not provided by our method. This can still be justified by a liberal choice of the selection method, e.g.\ in terms of a smaller probability in (R2). However, even with $90\%$ confidence in (R2), our simulations show very good coverage properties of our final segmentation. 
	
	\subsection{Hybridization}\label{sec:hybrid}
	Suppose that the set $\widetilde{\mathcal{B}}$ is a (possibly complex and highly redundant) system of sets satisfying (R2). The hybrid procedure is based on the important yet simple observation that (R2) remains valid if sets from $\widetilde{\mathcal{B}}$ are removed or enlarged. Therefore, to reduce the complexity (and redundancy) of $\widetilde{\mathcal{B}}$, we first neglect all sets that completely contain smaller sets. This might cause a loss of information, but on the other hand we are mostly interested in smaller sets as they contain the highest spatial information. This step yields a system $\mathcal B\subset \widetilde{\mathcal{B}}$ of sets still obeying (R2).
	
	In the following we describe how the system $\mathcal B$ can be hybridized with a segmentation $\mathcal W$ (i.e., a system of disjoint connected subsets of $\left[n\right]\times[n]$) such that on the one hand, the segmentation property is obeyed, and on the other hand (R2) is kept valid. There is a lot of freedom in choosing the segmentation algorithm, which allows the user to apply any method of choice and hence to generate a system of RoIs consisting of more naturally shaped segments. Possible examples include the famous Watershed segmentation algorithm (see below for a brief description), $k$-means clustering, or more recent AI-based techniques. \citet{SpatialHotspots} provide a comprehensive review on spatial hot-spot detection methodology. 
	
\begin{figure}[!b]
	\begin{tabular}{m{8cm}|m{2cm}m{1.4cm}m{2.2cm}}
		\toprule
		\textit{A segment $W$ containing a box $B$ is already valid and kept.} &\includegraphics[scale=1.3]{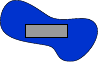}&\includegraphics[scale=1]{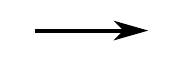}& \includegraphics[scale=1.3]{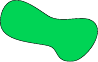}\\
		\midrule
		\textit{A segment $W$ intersecting with no box $B$ is dropped.} &\includegraphics[scale=1.3]{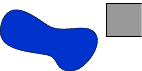}&\includegraphics[scale=1]{arrow2.pdf}& \\
		\midrule
		\textit{A segment $W$ not containing a complete box $B$ is merged with one.} &\includegraphics[scale=1.3]{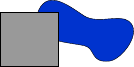}&\includegraphics[scale=1]{arrow2.pdf}& \includegraphics[scale=1.3]{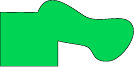}\\
		\midrule
		\textit{	If there is more than one box $B$ to validate a segment $W$, the one yielding the smallest resulting segment is chosen.} &\includegraphics[scale=1.3]{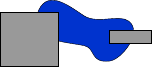}&\includegraphics[scale=1]{arrow2.pdf}&\includegraphics[scale=1.3]{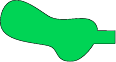} \\
		\midrule
		\textit{If the only possible validating box $B$ intersects with another segment $W'$, then the resulting regions is $R := W \cup B \cup W'$.} &\includegraphics[scale=1.3]{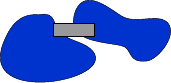}&\includegraphics[scale=1]{arrow2.pdf}&\includegraphics[scale=1.3]{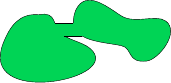} \\
		\midrule
		\textit{	If, in the above situation, another box might validate $W$, but another $W'$ cannot be validated without merging $W$ and $W'$, then we optimize this procedure such that the area is as small as possible. }  & \includegraphics[scale=1.3]{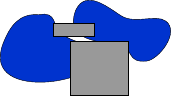}&\includegraphics[scale=1]{arrow2.pdf}&\includegraphics[scale=1.3]{r5.pdf}\\
		\bottomrule
	\end{tabular}
	\caption{\footnotesize Illustration of the hybridization algorithm: boxes $B$ (gray), segments $W$ to be validated (blue), final segments (green). }
	\label{tab:hybridization}
\end{figure}

	In the hybridization step we try to validate each segment $W \in \mathcal W$ using one (or more) of the sets $B \in \mathcal{B}$. Thus let $W \in \mathcal W$.  If there exists a $B \in \mathcal B$ such that $B \subset W$, then $W$ is already valid (in the sense of (R2)). If no such $B$ exists, then we merge $W$ with one of the intersecting sets to generate an enlarged valid segment. This step does, however, cause complications, as the chosen set $\hat B \in \mathcal B$ might intersect with other segments in $\mathcal W$. This issue is resolved as follows: 
	\begin{enumerate}
		\item We generate a set of 
		\textit{validation sets} $\mathcal B_{W} := \{ B \in \mathcal B \mid W \cap B \neq \emptyset \}$ for each $W \in \mathcal W$. 
		\item If $\mathcal B_{W} = \emptyset$, then $W$ has to be dropped. 
		\item If $\mathcal B_{W} \neq \emptyset$,  we check for each $B \in \mathcal B_W$ if $B \in \mathcal B_{W'}$ for another $W' \in \mathcal W$. This yields a list $\{B \in \mathcal B_W \mid \nexists W' \in \mathcal W, W' \neq W \text{ s.th. } B \in \mathcal B_{W'} \}$ of candidates for validating $W$. Every box $B$ in this list of candidates can now be used only to validate $W$ (and not to validate any other segment).
		\item From the list, we choose the one $\hat B$ yielding the smallest new valid segment $R = \hat B \cup W$. 
		\item If no such $B$ is found, we merge $W$ with another segment $W' \in \mathcal W$, which is again done such that the new resulting segment is as small as possible.
	\end{enumerate}
	The process described above is repeated iteratively until all segments have been validated or dropped, i.e.\ until (R3) is satisfied. The final, hybridized segmentation is then given as a set of regions $R$ denoted by $\widehat{\mathcal{ROI}}$. 
	We illustrate this procedure in Figure~\ref{tab:hybridization}.

	The next theorem guarantees that the hybrid selection $\widehat{\mathcal{ROI}}$ provides a valid segmentation in the sense of (R2) and (R3).

\begin{thm}\label{thm:seg}
	Let $\widehat{\mathcal{ROI}}$ be the regions of interest arising from the above hybridization algorithm of two systems of subsets $\mathcal B, \mathcal W \subset 2^{\left[n\right]\times\left[n\right]}$. Then the following holds true:
	\begin{enumerate}
		\item Let $\alpha \in \left(0,1\right)$. If the system of sets $\mathcal B$ obeys property (R2) in the sense of \eqref{eq:FWER} below, then also $\widehat{\mathcal{ROI}}$ does.
		\item If $\mathcal W$ is a segmentation, i.e., if the sets in $\mathcal W$ are pairwise disjoint, then also $\widehat{\mathcal{ROI}}$ is a segmentation.
	\end{enumerate}
\end{thm}

Clearly, the question whether (R1) and (R4)--(R5) are also satisfied, depends on the choice of $\widetilde{\mathcal{B}}$ and $\mathcal W$. If those are generated suitably, then this will be the case as discussed in the following section.

\subsection{Practical implementation}\label{sec:MISCAT}

In the following we will focus on two specific methods to generate the sets $\widetilde{\mathcal{B}}$ and $\mathcal W$, which we used in our implementation. To generate $\widetilde{\mathcal{B}}$, we employ the MISCAT procedure introduced in \citet{pwm16}, which we will briefly describe in the following for convenience (see also \citet{mplw20} for a comprehensive review of the MISCAT procedure).
We start our segmentation by determining rectangular regions of interest via a multiple testing approach. The rectangular regions will be referred to as boxes, denoted by $B_{\bx,\bh}$. The subscript $\bx$ denotes the position of the upper left corner of the box within the grid of pixels, while the subscript $\bh=(h_1,h_2)$ denotes the side lengths of the box. For a given $\bh$ we consider all $B_{\bx,\bh}\subset\{\bx_\bi\,|\,\bi\in\,[n]\times[n], \bi + \bh \in\,[n]\times[n]\}$ and furthermore we use several different  \textit{scales} $\bh$ ranging from small to large, so that all together we consider a highly redundant system of boxes that guarantees very good detection properties of the MISCAT method. To test whether a box $B_{\bx,\bh}$ contains markers, we design pairs of functions $\varphi_{\bx,\bh}, \Phi_{\bx,\bh}$ such that
\begin{align}\label{eq:probe}
	\left\langle f \ast h, \Phi_{\bx,\bh}\right\rangle  = \left\langle f, \varphi_{\bx,\bh}\right\rangle
\end{align}
with the fluorescence intensity $f = f_Nf_p$ where $\ast$ denotes spatial convolution. The function pairs $\varphi_{\bx,\bh}, \Phi_{\bx,\bh}$ will in practice be generated by a kernel somewhat similar to Wavelets, such that they form a  multiscale system that adapts to the PSF $\psf$. This ensures optimal detection power of MISCAT by fine-tuning the kernel, see \citet{pwm16}. Given equation \eqref{eq:probe} and having in mind that $\mathbf Y^{\text{S},1} = (Y_{\bi}^{\text{S},1})_{\bi\in[n]\times[n]}$ consists of the most highly resolved measurements available, $\left\langle\mathbf Y^{\text{S},1}, \Phi_{\bx,\bh}\right\rangle $ can serve as a \textit{local test statistic} for testing whether $f_{|_{B_{\bx,\bh}}} \not\equiv 0$.  The local test statistics are then combined by taking the maximum after subtracting a scale-dependent penalization (to ensure equal contribution of the differently sized boxes). This yields a test statistic for a multiple test over all considered boxes with a controlled family wise error rate as a consequence of Theorem 4 in \citet{pwm16}. It is furthermore shown in Theorem 1 of \citet{pwm16}, that quantiles of this maximum (or scan) statistic can be simulated using a suitable Gaussian approximation, which gives then rise to local (taking a suitable scale-penalization into account) critical values  $c_{\bh,\alpha}$. The outcome of this first step is then a set $\widetilde{\mathcal{B}}$ of candidate RoIs given by
\begin{center}
	$\widetilde{\mathcal{B}}=\{B_{\bx,\bh}\,|\,\left\langle  \mathbf Y^{\text{S},1},\Phi_{B_{\bx,\bh}}\right\rangle\geq c_{\bh,\alpha}\rangle\}.$
\end{center}
Note, that the resulting candidate RoIs are boxes of different sizes $\bh$. The MISCAT procedure at level $\alpha$ is constructed in such a way that
\begin{equation}\label{eq:wFWER}
	\mathbb P \left[\text{There is a pair } (\bx,\bh): B_{\bx,\bh}\in\widetilde{\mathcal{B}} \,|\,\text{ no markers at all}\right] \leq\alpha.
\end{equation}
Furthermore, as the distribution of the maxima of the statistics $\left\langle  \mathbf Y^{\text{S},1},\Phi_{B_{\bx,\bh}}\right\rangle$ does not depend on the actual positions of the $N$ markers in the image, we have \textit{subset pivotality}, cf. \citet{westfall1993resampling}. 
As a consequence, the FWER control \eqref{eq:wFWER} for MISCAT implies that with high probability \textit{all selected boxes contain a marker}:
\begin{equation}\label{eq:FWER}
	\inf_{C\in\mathcal{C}_N}\mathbb P \left[f_{|_{B_{\bx,\bh}}} \not\equiv 0 \text{ for all } B_{\bx,\bh} \in \widetilde{\mathcal{B}}\,|\,\text{configuration }C\right] \geq 1-\alpha,
\end{equation}
where $\mathcal C_N$ is the set of all possible configurations (distributions) of $N$ markers in the image of interest.
This is a FWER control over the selected system of boxes in the strong sense, which in general -- more precisely without subset pivotality -- does not follow from the weaker FWER control \eqref{eq:wFWER}.

For the generation of the segmentation $\mathcal W$, we use here the well-known watershed algorithm, whose name is in reference to a geological watershed, which separates adjacent drainage basins. The algorithm can be applied to grey scale images, which are interpreted as a topographic map, with the brightness of each point representing its height. The algorithm  finds the lines that run along the tops of ridges (see \citet{m94} for more details).

As a corollary of Theorem~\ref{thm:seg}, we obtain for the particular choice of MISCAT at level $\alpha$ and the watershed segmentation $\mathcal W$ with any choice of tuning parameter asymptotically
\begin{center}$
	\inf_{C\in\mathcal{C}_N}\mathbb{P}\left[\text{Each } \widehat R_i\in \widehat{\mathcal{ROI}}\text{ contains } \geq 1 \text{ marker}\,|\,\text{config. }C\right]\geq1-\alpha.$
\end{center}
This means that the so obtained hybrid segmentation inherits the strong control of the FWER from the MISCAT procedure. 

\begin{figure}[h]
		\centering
		\includegraphics[width = 0.9\textwidth]{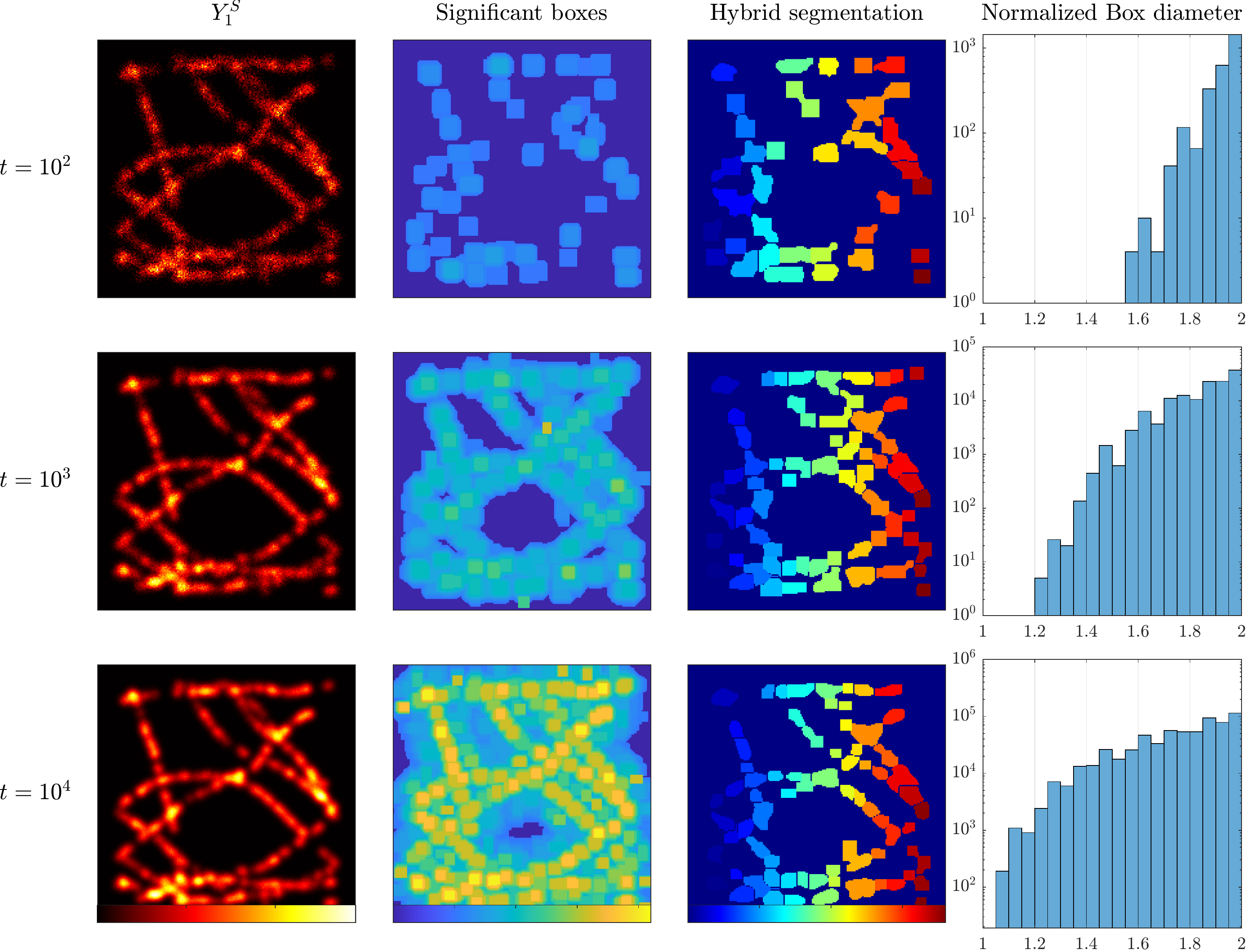}
		\caption{From left to right: 1.) Simulation of filamentous structures: Data sets for different numbers $t$ of excitation light pulses. 2.) MISCAT segmentation where boxes are shown in a color-scale inverse to box area and smaller boxes are drawn on top of larger boxes. 3.) Hybrid segmentation. 4.) Histograms of the segment diameter normalized by the FWHM of the PSF ($\sqrt{|R|}/FWHM$).}
		\label{fig:segmentation}
	\end{figure}

	We illustrate the performance of this hybrid segmentation procedure in Figure \ref{fig:segmentation}.  The left panel shows simulated data for artificial filamentous structures (resembling among other things the cytoskeleton in cells) generated by the convolution model \eqref{eq:conv} for a STED microscope with different numbers  $t$ of excitation pulses. The second panel depicts all boxes selected as significant by the MISCAT procedure color-coded by size. The MISCAT test is able to find a set of significant boxes of various sizes at controlled FWER. However, many of  these rectangles are ''too large'' in the sense that they do not provide spatial information of markers with sufficient accuracy. Moreover, the set $\widetilde{\mathcal{B}}$ is highly redundant, which is typical for systems satisfying (R2). The third panel of Figure \ref{fig:segmentation} shows the result of the suggested hybridization, which covers reasonably well the whole structure with non-overlapping significant segments. The fourth panel depicts a histogram of the normalized segment diameter. In total we conclude that all goals (R1)--(R5) are met in general. (R2) and (R3) are ensured by Theorem \ref{thm:seg} as discussed above independent of the excitation time $t > 0$. The other requirements clearly depend on the data quality via $t$. For $t = 1000$ (which is an experimentally reasonable parameter), both (R1) and (R4) are visually satisfied. The shape of the RoIs is mostly determined by the watershed algorithm, and we find that this meets also (R5). The coverage of the structure as well as the number of found segments increases with increased brightness of the image, while the average size of the found significant boxes and hybrid segments decreases.
	
	\section{Statistical molecule counting}\label{sec:estimation}

In this section we discuss how to estimate the local number of molecules inside a region $R$ from the data. Furthermore, we provide a central limit theorem (CLT) for our estimator, which immediately allows us to construct (asymptotic) confidence intervals quantifying the precision of our counting method. Later on we will apply the methodology presented here to all regions in the previously discussed hybrid segmentation  $\widehat{\mathcal{ROI}}$. 

\bigskip

For the derivation of our estimator, suppose for a moment that all the $N$ molecules of the specimen are clustered at a single point, this is $\psi(j) = \mathbf k \in \left[n\right]\times\left[n\right]$ for all $1 \leq j \leq N$ with some fixed $\mathbf k \in \left[n\right]\times \left[n\right]$. In this case it follows immediately from the definition that
\[
\sum_{\bi \in \left[n\right]\times\left[n\right]}s_l \left(\bx_{\bi}\right) = \sum_{\bi \in \left[n\right]\times\left[n\right]} \sum_{j=1}^N \brightx_j(\bx_{\bi})^l  = \sum_{\bi \in \left[n\right]\times\left[n\right]} \sum_{j=1}^N \bright_j^l\cdot\psf(\bx_{\bi}-\bx_{\mathbf k})^l.
\]
In the following we assume that the brightness $f_p$ is a function of the location only, and not a function of the individual molecules and we further assume a certain regularity of the function $f_p$ such that it can be assumed constant within sufficiently small regions. 
As the brightness depends mostly on local parameters such as the pH, this is a realistic restriction of our original model where each marker was allowed to have an individual brightness. We obtain
\[
\sum_{\bi \in \left[n\right]\times\left[n\right]}s_l \left(\bx_{\bi}\right) = \bright^l N \sum_{\bi \in \left[n\right]\times\left[n\right]}\psf(\bx_{\bi}-\bx_{\mathbf k})^l.
\]
In all practically relevant examples, the PSF $\psf$ is rapidly decaying and can hence be considered as periodic, which justifies the additional assumption that
\begin{align}\label{eq:Hl}
	H_l := \sum_{\bi \in \left[n\right]\times\left[n\right]}\psf(\bx_{\bi}-\bx_{\mathbf k})^l
\end{align}
is independent of the location $\mathbf k \in \left[n\right]\times \left[n\right]$. Consequently, it holds
\begin{align}\label{eq:N}
	N=\frac{H_2}{H_1^2}\frac{\left(\sum_{\bi\in \left[n\right]\times\left[n\right]}s_1\left(\bx_\bi\right)\right)^2}{\sum_{\bi\in \left[n\right]\times\left[n\right]}s_2\left(\bx_{\bi}\right)}.
\end{align}
In the situation of Theorem~\ref{Le:trans}, the quantities $\bs\left(\bx_\bi\right)$ can be immediately estimated, as then
\begin{align}\label{eq:se}
	\bs(\bx_\bi)=(\trans^{-1}(\bD(\bx_\bi))+O\Big(\gamma(\bx_\bi)^{\md+1}\Big).
\end{align}
This leads to a natural plugin estimator for the total number of molecules, where $\bD(\bx_\bi)$ is estimated by the relative frequencies of detecting $0,1,\ldots,\md$ photons at position $\bx_\bi$.

\bigskip

Let us now return to the general situation that we want to estimate
\begin{center}$
	N_{R}=\#\{\text{markers in region } R\} = \int_R f_N(x) \,\mathrm d x$
\end{center}
for some region $R \subset \left[0,1\right]^2$. For this, we pose the following physical assumption:
\begin{ass}\label{ass:bright}
	The brightness within the region $R$ is approximately constant if $R$ is sufficiently small and the functions $H_l$ defined in \eqref{eq:Hl} do not depend on $\mathbf k$.
\end{ass}
This assumption is (approximately) valid for many experimental settings as long as $R$ is not too large, as the brightness depends mostly on local conditions such as temperature, pH value and so on. 
Under Assumption~\ref{ass:bright} it seems now natural to replace the sum over all locations $\bx_\bi$ in \eqref{eq:N} by the sum over all $\bx_{\bi} \in R_\varepsilon$ with a slightly enlarged segment $R_\varepsilon = \left\{x \in \left[0,1\right] ~\big|~ \text{dist}(x,R) \leq \varepsilon\right\}$. The rationale behind is that the rapid decay of the PSF $\psf$ ensures that molecules inside $R$ will only or at least mostly contribute to those $\bx_\bi \in R$. In practice we choose $\varepsilon$ of the order of the FWHM and enlarge only as long as $R_\varepsilon$ does not intersect with any other enlarged segment. Let us introduce $\mathcal I_{R_\varepsilon} := \left\{\bi \in \left[n\right]\times\left[n\right]~\big|~\bx_\bi \in R_\varepsilon\right\}$. Then the above considerations give rise to the plugin estimator
\begin{align*}
	\widehat{N}_R=\frac{H_2}{H_1^2}\frac{\left(\sum_{\bi\in \mathcal{I}_{R_\varepsilon}}\hat s_1\left(\bx_\bi\right)\right)^2}{\sum_{\bi\in \mathcal{I}_{R_\varepsilon}}\hat s_2\left(\bx_{\bi}\right)}\quad\text{with}\quad \hat\bs(\bx_\bi)=\trans^{-1}\left(Y_{\bi}^{\text{C},0},...,Y_{\bi}^{\text{C},\md}\right)/t.
\end{align*}
Note that the inversion of $\trans$ might introduce uncertainties (and actually does, cf. Figure~\ref{fig:preprocessing-histograms} in our numerical simulations), and as the noise level of $\hat s_k(\bx_\bi)$ increases geometrically with $k$ in view of the $f_p^k$-dependency, we decided to use only $\hat s_1(\bx_\bi)$ and $\hat s_2(\bx_\bi)$ to infer on $N_R$. Nevertheless, it is in principle possible to improve the estimate for $N_R$ based on higher order contributions $s_k(\bx_\bi)$, $k \geq 2$ along the above considerations.

The following central limit theorem for the estimator $\widehat{N}_R$ is based on asymptotics for $t\to\infty$ subject to $\md=\md(t)\to \infty$. Practically, it is clearly unrealistic to build microscopes with more and more detectors, but for asymptotic considerations such an assumption is unavoidable, as for $t \to \infty$ arbitrarily high photon coincidences will occur and need to be recorded appropriately. This has already been observed in the model bias \eqref{eq:modelbias}, and thus by $\md=\md(t)\to \infty$ we can assume that $t$ is large enough such that the model bias is irrelevant. Since the influence of the size of $\md$ is of practical interest, finite sample bounds of Berry-Esseen type with respect to $t$ as well as $\md$ are additionally provided. Practically, we will show in our simulation study, that $\md = 4$ for suffices for up to $N = 40$ molecules, $\md =6$ for up to $N=100$ molecules, and $\md = 8$ even for around $N = 150$ molecules per segment. So only for extremely dense objects, an improvement of the experimental setup might be necessary.

Denote by $\Pi^d_{(k_1,\ldots,k_l)}:\mathbb{R}^d\to\mathbb{R}^{l}$ the projection of a vector in $\mathbb{R}^d$ onto the vector of its coordinates $k_1,\ldots,k_l$.

\begin{thm}\label{thm:uglythm}
	Let $R\subset[0,1]^2$ and let $\md=\md(t)\to \infty$, such that $\md\leq5\log(t),$ Suppose that the region $R$ contains at least one fluorescent marker, that Assumption~\ref{ass:bright} is satisfied, and that any marker in $R$ has individual brightness smaller than $0.5$ (see Remark~\ref{rem:uglythm} (v)). 
	Let
	\begin{center}$
		\Sigma_{\mathbf{E}}(\bx_\bi)=
		\left(
		\begin{cases}
			E_j(\bx_\bi)(1-E_j(\bx_\bi)) & j=k\\
			-E_j(\bx_\bi)E_k(\bx_\bi) & j\neq k
		\end{cases}
		\right)_{j,k=1}^{\md},$
	\end{center}
	and consider an arbitrary but fixed ordering $\bx_{\bi_1},\ldots,\bx_{\bi_{|R|}}$ of the points $\bx_\bi\in R$. Further, define the block diagonal matrix $\Sigma_R$ by
	$
	\mathrm{diag}\left(\Sigma_{\mathbf{E}}(\bx_{\bi_j})\right)_{j=1,\ldots,|R|}
	$
	and let $\transclt:\mathbb{R}^{\md|R|}\to\mathbb{R}$ be given by
	\begin{center}$
		\transclt\left(\mathbf{y}\right) := \frac{H_2}{H_1^2}\cdot\frac{\left\langle\mathbf{1}_{\md|R|},\left(\Pi^{\md}_1\trans^{-1}(y_{j\md+1},\ldots,y_{(j+1)\md})^T\right)_{j=1,\ldots,|R|}\right\rangle^2}{\left\langle\mathbf{1}_{\md|R|},\left(\Pi^{\md}_2\trans^{-1}(y_{j\md+1},\ldots,y_{(j+1)\md})^T\right)_{j=1,\ldots,|R|}\right\rangle}.$
	\end{center}
	Assume that the Hessian matrix of $\transclt$, $\mathrm{Hess}_{\transclt}$, exists and is bounded in a neighbourhood of $\mathcal{E}$, and that
	\begin{align}\label{eq:sigmaR}
		\sigma_R^2=\nabla\transcltD^T\Sigma_R\nabla\transcltD>0,
	\end{align}
	where $\mathcal{E}=(\mathbf{E}(\bx_{\bi_j}))_{j=1,\ldots,|R|}$.
	Let $Z$ be a random variable following a centered normal distribution with variance $\sigma_R^2$. If $t$ is sufficiently large, there exists a constant $C>0$ such that
	\begin{align}\label{eq:clt}
		\sup_{s\in\mathbb{R}}\left|\mathbb{P}\left(\sqrt{t}\Big(\widehat N_R-N_R\Big)\leq s\right)-\mathbb{P}\Big(Z\leq s\Big)\right|\leq C\frac{\log(t)^{\frac{2}{3}}}{t^{\frac{1}{6}}},\qquad\text{as}\qquad t \to \infty.
	\end{align}
	
\end{thm}

\begin{rem}\label{rem:uglythm}
	\begin{itemize}
		
		\item[(i)] In particular, the above theorem ensures that
		\begin{center}$
			\sqrt{t}\Big(\widehat N_R-N_R\Big)\stackrel{\mathcal{D}}{\longrightarrow}\mathcal{N}(0,\sigma_R^2),\qquad\text{as}\qquad t \to \infty,$
		\end{center}
		where $\stackrel{\mathcal{D}}{\longrightarrow}$ denotes convergence in distribution.
		\item[(ii)] Recall that the asymptotic considerations are with respect to the number $t$ of laser pulses. As $t$ tends to infinity, the number of pixels remains fixed. Therefore, the collection of regions $R$ considered does not change asymptotically and contains only finitely many elements. Therefore, naturally, \eqref{eq:clt} holds uniformly in $R$.
		\item[(iii)] We let $\md\to\infty$ as $t\to\infty$, as for a fixed number $\md$ the model bias \eqref{eq:gamma} may otherwise asymptotically dominate the quantity $\sqrt{t}(\widehat N_R-N_R)$. The upper bound on the speed of convergence of $\md$ is needed in the proof, as with $\md$ the number of events of the multinomial distribution tends to infinity. However, this restriction is irrelevant in practice, where the number of detectors that can be realized is limited.
		\item[(iv)] In this work, the focus is on the estimation of local numbers of molecules, $N_R$. However, similar to \eqref{eq:N}, we immediately obtain an expression for $p_R$, the brightness in a region $R$, as
		$
		p_R=H_1\sum_{\bi\in \mathcal{I}_R}s_2\left(\bx_\bi\right)/(H_2\sum_{\bi\in \mathcal{I}_R}s_1\left(\bx_{\bi}\right)),
		$
		giving rise to a plug-in estimator $\hat p_R$. The theoretical properties of $\hat p_R$ can be analyzed with the same techniques as applied in the analysis of $\widehat N_R$.
		\item[(v)] We assumed that any marker in $R$ has individual brightness smaller than $0.5$. This assumption is needed for technical reasons but it is not restrictive, as typical brightness values are of the order of $0.02$, as also used in our simulations. A brightness of 0.5 would entail that a marker emits a photon on average once every two pulses, which is unrealistic.
		\item[(vi)] Given that $\md\to\infty$, the model bias vanishes asymptotically. 
		Our numerical simulations show that as few as 6 detectors suffice to deal with local numbers of molecules of about $100$ (see Figure~\ref{fig:single-cluster-vs-md}). The experimental setup we used (as shown in Figure~\ref{fig:experimental_setup}) exploits $\md = 4$ detectors, and we find from Figure~\ref{fig:single-cluster-vs-md} 
		that it is expected to be accurate to treat local molecule numbers of around $30-40$.
	\end{itemize}
\end{rem}
Theorem~\ref{thm:uglythm} gives rise to asymptotic confidence intervals for $N_R$: 
$
\left[\widehat N_R-\frac{\hat\sigma_R}{\sqrt t},\widehat N_R+\frac{\hat\sigma_R}{\sqrt t}\right],
$
where $\hat\sigma_R$ is the plug in estimator for the asymptotic standard deviation $\sigma_R$ defined in \eqref{eq:sigmaR}, that is, $\hat\sigma_R=\nabla\transclt(\widehat{\mathcal{E}})^T\Sigma_R\nabla\transclt(\widehat{\mathcal{E}})$, with $\widehat{\mathcal{E}}=1/t\cdot(Y^{\text{C},0}_{\bi},...,Y^{\text{C},\md}_{\bi})_{\bi\in R}$.


\section{Numerical study}\label{sec:numerical}

In this section we will investigate the finite sample properties of our estimation procedure and the overall algorithm.  The complete MATLAB\textsuperscript{\textregistered} code including all examples discussed in the manuscript at hand is available under
	\begin{center}
		\small \url{https://github.com/jkfindeisen/antibunching_microscopy_analysis_2022}.    
	\end{center}
	
\begin{figure}[!tb]
	\centering
	\includegraphics[width = 0.9\textwidth]{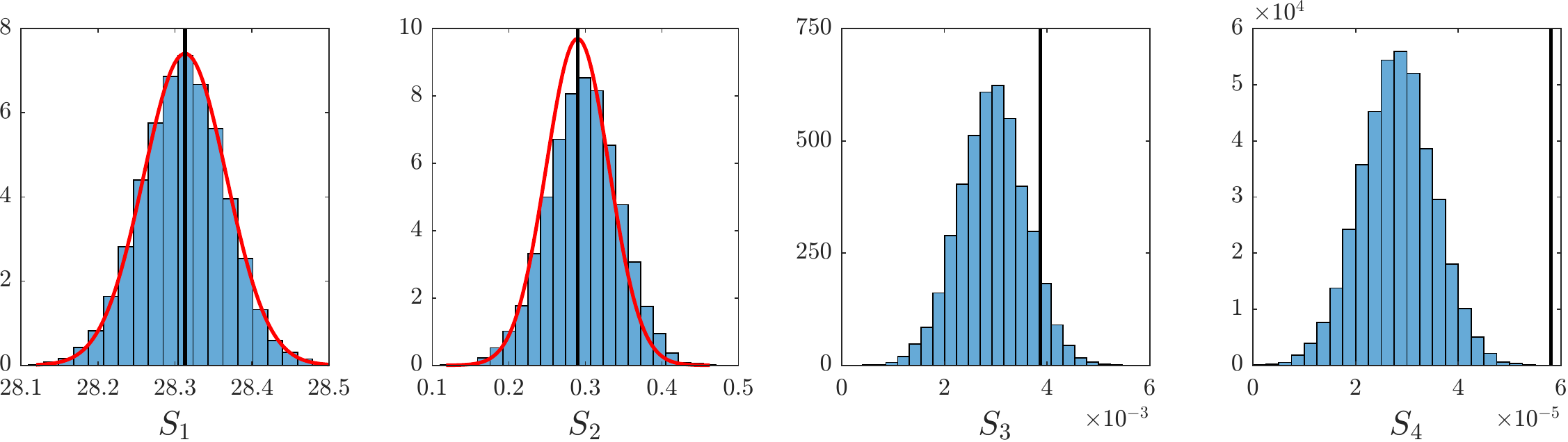}
	\caption{Estimation of $S_l:= \sum_{\bi \in \left[n\right]\times \left[n\right]} s_l\left(\bx_{\bi}\right)$ for simulations of single, isolated cluster of $N=20$ molecules with brightness $p=0.02$ and a Gaussian PSF of FWHM=4px and $t=10^4$ pulses per pixel. Histograms of estimated $S_i$ for many repetitions. Vertical lines (black) present the true mean values. Red curves represent normal distributions with true variances around the true means.} 
	\label{fig:preprocessing-histograms}
\end{figure}

	As a first step, we assumed a certain number of independent and identically behaving molecules with a fixed brightness $p$ to be located all at the same position and to be imaged with the antibunching microscope operating only on confocal mode. The recording of the molecule sample is performed by scanning on a square-lattice like grid relative to the molecules' position and the grid spacing is here defined relative to the resolution of the microscope, i.e. with a FWHM of the PSF given in scanning grid pixel sizes. For a single cluster, segmentation of the data was omitted and the whole measurement area was assumed to belong to the single segment representing the molecule cluster. In Figure~\ref{fig:preprocessing-histograms} we depict histograms of the empirical distributions of $S_l := \sum_{\bi \in \left[n\right]\times \left[n\right]} s_l\left(\bx_{\bi}\right)$ together with the theoretical expectations of $S_l$. It can be seen that for typical experimental conditions, the first two orders of $S_l$ are reasonably well distributed around their expectations and can be used for retrieving the number and brightness of the molecules. For higher orders, systematic deviations are visible in the distributions of $S_l$, most probably a side effect of the ill-conditioned inversion of $\trans$ under these circumstances. 
	
\begin{figure}[!tb]
	\centering
	\includegraphics[width = 0.9\textwidth]{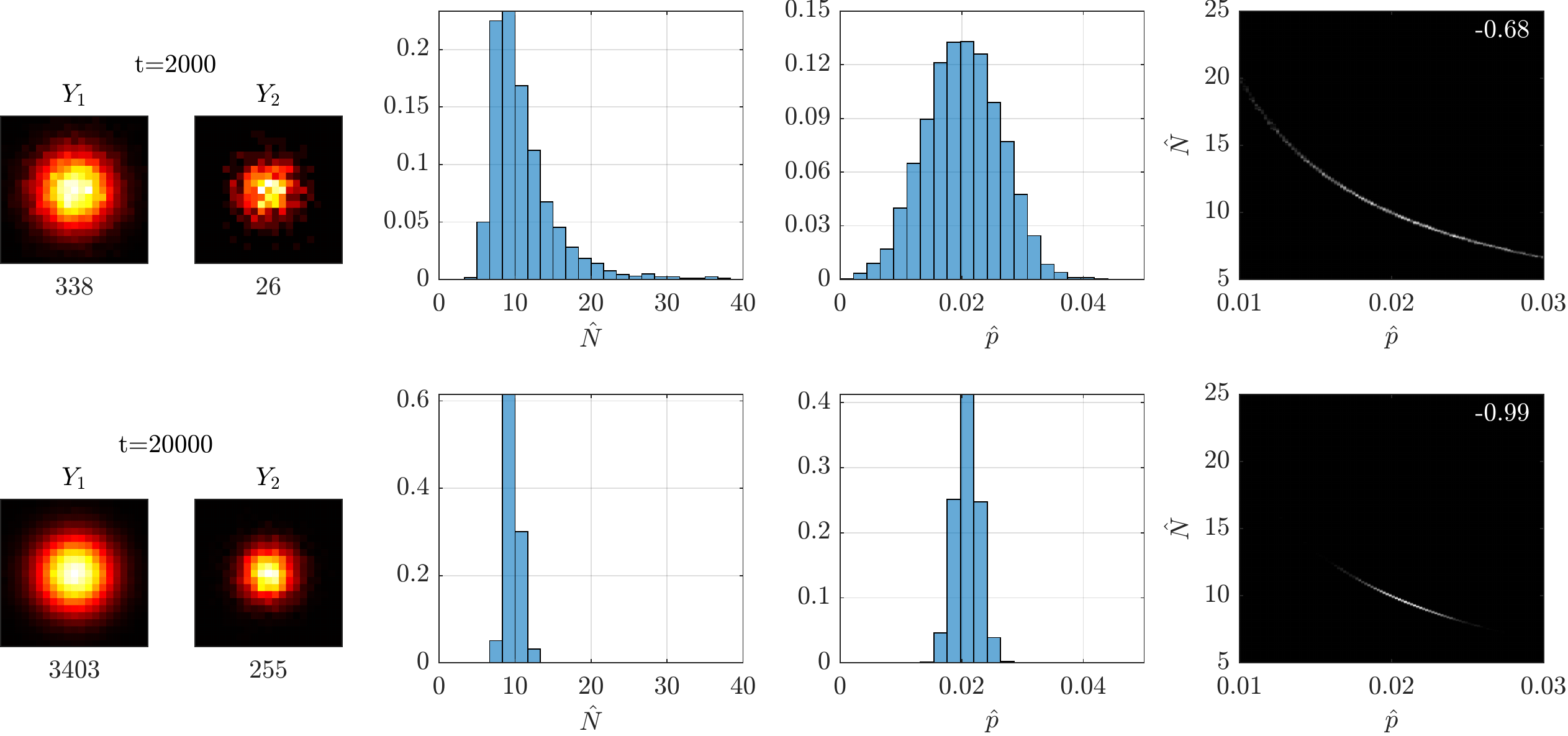}
	\caption{Histograms of estimated $N$ and $p$ for simulations of single, isolated cluster of $N=10$ molecules with brightness $p=0.02$ and a Gaussian PSF of FWHM=4px and $t=10^3$ and $t=10^4$ pulses per pixel, respectively.} 
	\label{fig:single-cluster-histograms}
\end{figure}

	This first simulation shows that it is possible to estimate the number of molecules in a single, isolated cluster from $S_1$ and $S_2$. This is further illustrated in Figure~\ref{fig:single-cluster-histograms}, where we investigate the final estimator for the total number of molecules $N$ (and the corresponding one for the common brightness $p$ as discussed in Remark \ref{rem:uglythm}(iv)) in a similar setting. Again we use a single cluster with $N = 10$ molecules with common brightness $p = 0.02$ and omit the segmentation step by assuming that the whole measurement area was assumed to belong to the single segment representing the molecule cluster. Repetition of the simulation results in distributions for $\hat N$ and $\hat p$, which are depicted as histograms. It should be noted that the estimated molecule numbers and brightnesses are correlated and the joint distribution is concentrated along a hyperbola in the number-brightness plane (see Fig.~\ref{fig:single-cluster-histograms}c), i.e. the product of the molecule number and its brightness is quite well estimated compared to the knowledge about the single factors. For larger numbers of illumination pulses the distributions of the estimated numbers and brightnesses becomes more concentrated, more symmetric and less biased, indicating that the employed estimators converge to the true underlying parameters. The precision is mainly limited by the number of repetitions, i.e. light pulses, that the sample can be illuminated with at every scan position. A few thousand excitation pulses per scan position have been reported to be possible without significant observable photobleaching of the molecules.

\begin{figure}[!tb]
	\centering
	\includegraphics[width = 0.9\textwidth]{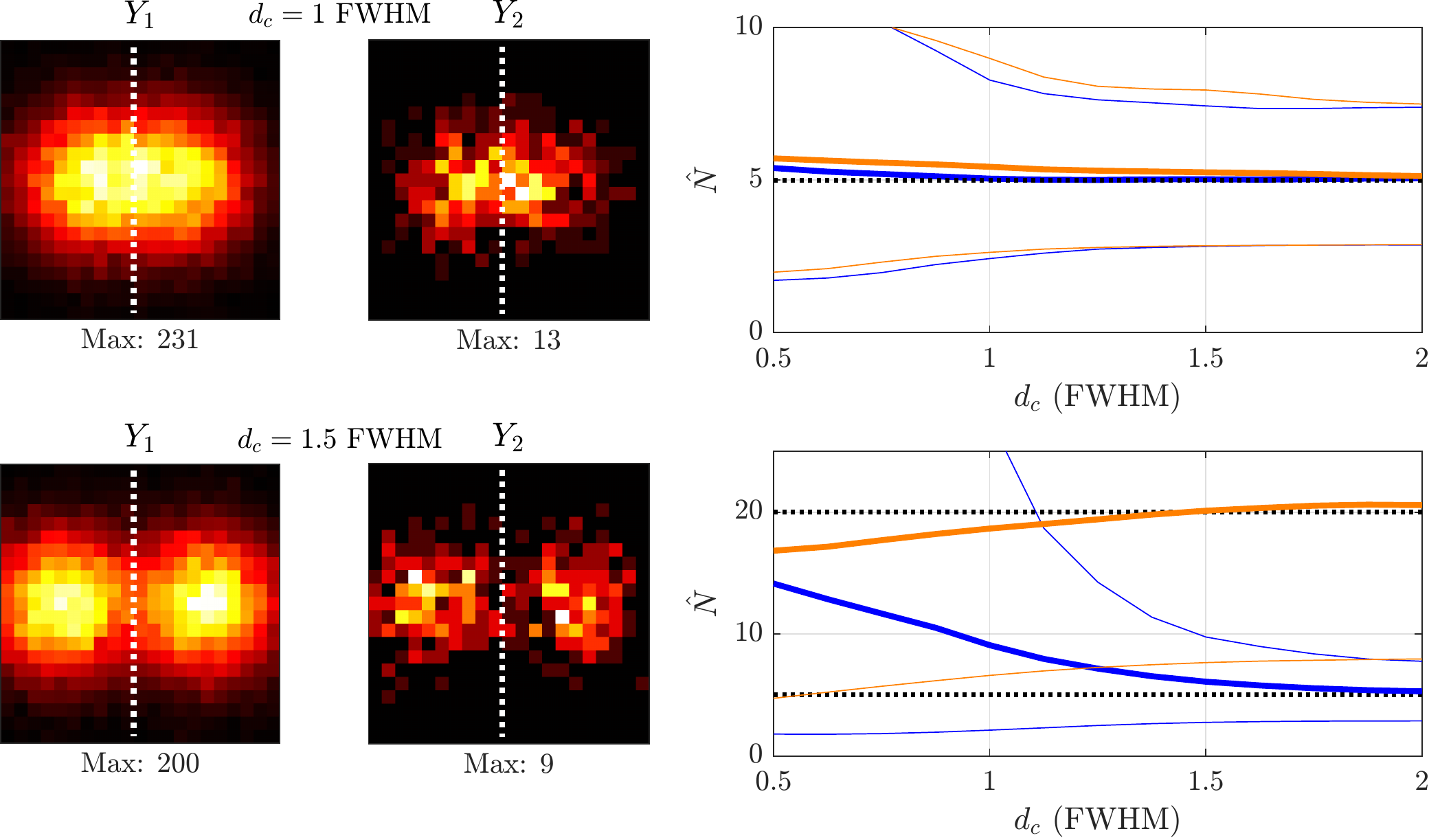}
	\caption{Estimated $N$ for simulations of two clusters of molecules with a defined distance $d_c$ (given in multiples of the FWHM) with brightness $p=0.02$, $t=3000$ and a Gaussian PSF of FWHM=4px. (left) Example images of two clusters with five molecules each. Dotted white line represents the segment border. (right) Mean estimated number of molecules for each cluster and mean lower and upper bounds of the confidence intervals (thin lines) in dependence of the distance $d_c$ between the clusters. Black dotted line represents the true number of molecules in each cluster. Top: two clusters with $N_1=N_2=5$ and bottom: $N_1=5, N_2=20$.} 
	\label{fig:two-clusters}
\end{figure}

	To study the influence of boundary effects, i.e. molecules residing close to the boundary of an image segment, we simulated two clusters of molecules with a defined distance that is on the order of the FWHM of the PSF. In that way the images of the two clusters are partly overlapping. Two segments were created such that they fully covered the whole simulated image space and the border between the two segments was localized in the middle between the clusters. The analysis in each segment will be compromised by photons that are located in the respective other segment.
The results depicted in Figure~\ref{fig:two-clusters} show that for a reasonable cluster distance (of approximately the FWHM of the PSF) the systematic boundary effects tend to become relatively small. It should be noted that our hybrid segmentation naturally tends to avoid large amounts of signal close to segment borders.
	
	\begin{figure}[!tb]
	\centering
	\includegraphics[width = 7cm]{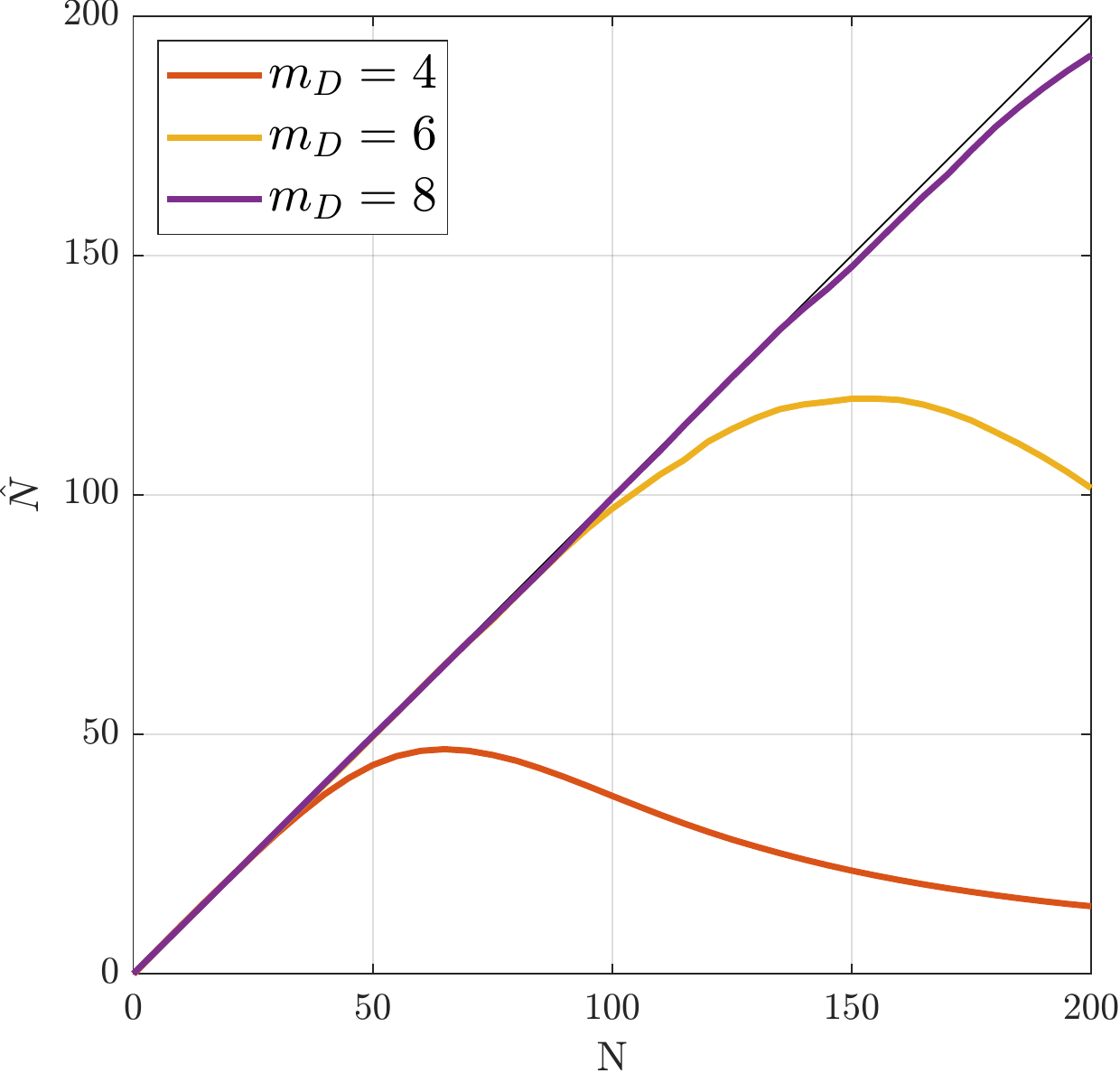}
	\caption{Median of estimated number of molecules for a simulation of a single cluster ($p=0.02$, $t=2\cdot10^4$) in dependence of the number of detectors used $m_d$. The visible bias is strongly reduced for larger $m_d$. All internal orders equal $m_d$. Graphs are slightly smoothed.} 
	\label{fig:single-cluster-vs-md}
\end{figure}
	
	Another source of systematic bias are very dense accumulations of molecules. Therefore, we performed simulations in which we increased the number of detectors. We found that the range of molecules that can be estimated without visible bias strongly increases already for a number of detectors that is increased only moderately. Four parallel detectors, which are currently used experimentally, allow to count up to 50 molecules per diffraction limited sample volume without large bias with our method. With eight parallel detectors this limit could be lifted to approximately 200, cf. the left panel in Figure~\ref{fig:single-cluster-vs-md}.

The overall performance of our algorithm is tested in a simulated arrangement of markers consisting of several clusters of molecules with varying brightness, cf. Figure~
\ref{fig:workflow}. It shows that the estimation of the local number of molecules is accurate in all segments, and that the FWER control \eqref{eq:FWER} is in fact kept. It must be said that some of the constructed confidence intervals seem to be rather large (indicating a small estimated value $\hat p$ in that region), which is, however, unavoidable when asking for strong error controls such as \eqref{eq:FWER}. In addition, in Figure~
\ref{fig:workflow}, the true values of $N_R$ are located on the boundary of the confidence intervals in multiple cases, indicating that any method with smaller confidence intervals, which are centered at $\widehat N_R$, would be susceptible to violations of the property \eqref{eq:coverage}.

\section{Counting molecules in a DNA origami measurement}\label{sec:origami}

\begin{figure}[!tb]
	\centering
	\includegraphics[width=0.9\textwidth]{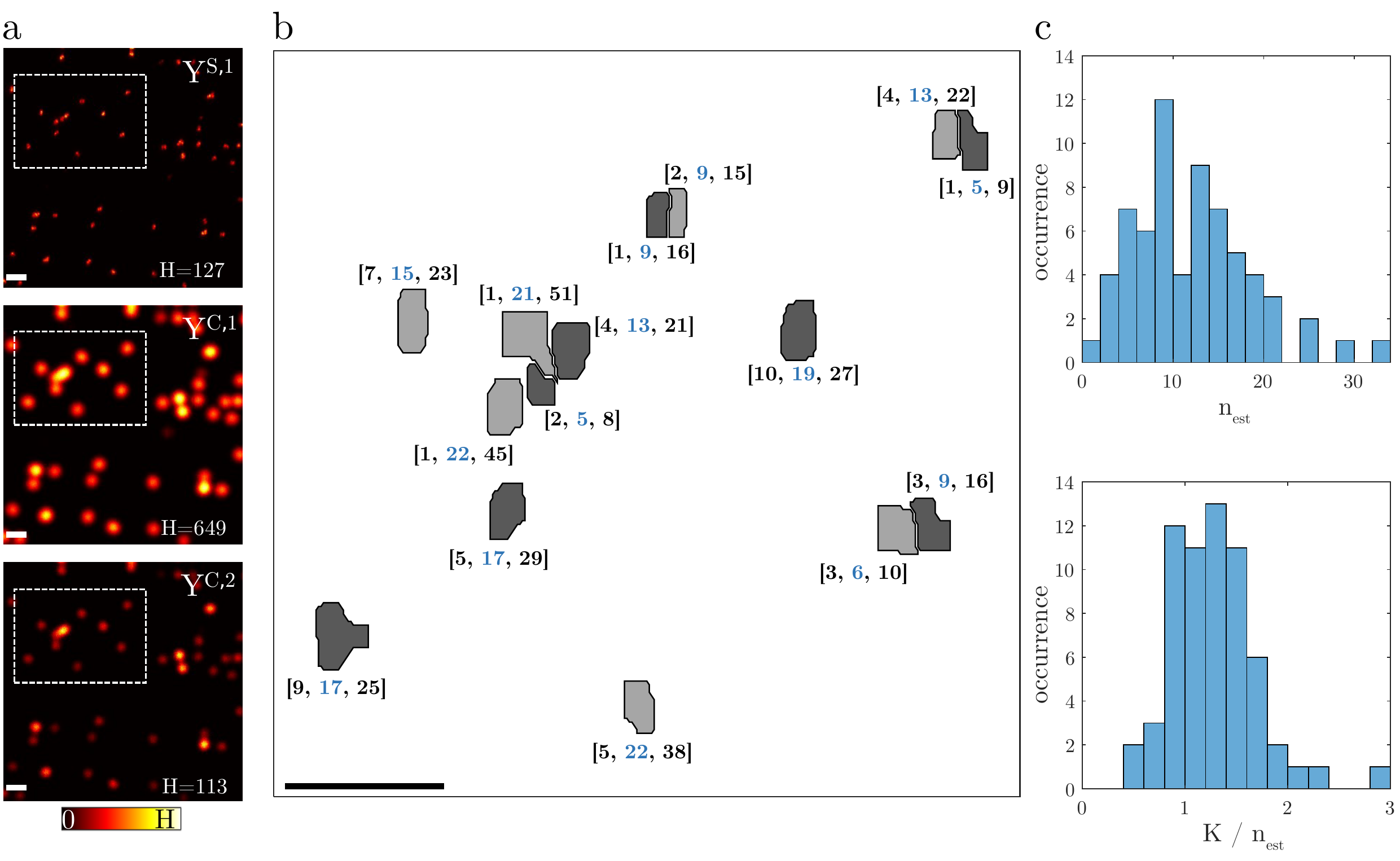}
	\caption{ \footnotesize Application to recordings of DNA origami sheet structures labelled with ATTO 647N molecules (for experimental details see \citet{t15}). The structures were immobilized at a low concentration on a glass surface and measured with confocal and STED microscopy. Each DNA origami can accommodate up to 24 fluorophores (12 in a line). \textbf{a} STED (top) and confocal (number of pulses per pixel t=2170, pixel size 10 nm) one-photon (center) and two-photon (bottom) detection images. \textbf{b} Segmentation, estimated number of fluorophores in each segment and confidence intervals for the data in the dotted rectangle. \textbf{c} Histogram of estimated number of fluorophores in each segment of the recorded data (top) and histogram of the width of the confidence interval ($K$) divided by the estimated number of fluorophores (bottom). Scale bars, 500 nm.}
	\label{fig:origami-measurement}
\end{figure}

To establish the validity of the suggested confidence intervals on real data we used DNA origami sheets, which is an artificial structure that allows to attach a relatively well defined number of molecules at defined positions within a diffraction-limited volume (\citet{schmied12}). The designed DNA origami sheet contained up to 24 fluorophores arranged in two lines of up to 12 fluorophore binding sites and the distance between the two lines was $\mathord{\sim} 70$ nm, which is not resolvable in confocal mode of the microscope. Due to imperfections in the folding efficiency of DNA, the expected number of fluorophores per DNA origami was only $\mathord{\sim} 19$ (\citet{t15}). Confocal microscopy cannot spatially resolve the single lines of fluorophores, but it allows to obtain a sufficient statistic on one and two photon detection events. A subsequent STED recording with improved lateral resolution (fivefold over confocal microscopy) resolved the molecular distribution within a single DNA origami sheet. Further experimental details are laid out in \citet{t15}. Here, we re-analyzed the recorded data to yield estimates and confidence intervals on the number of fluorophores in either resolved DNA origami sheets or even single lines in these sheets (see Fig.~\ref{fig:origami-measurement}). The division of the data into suitable segments using our hybrid MISCAT and watershed segmentation approach was performed on the high resolution STED data, while the molecular number and brightness was estimated from the less well resolved but much brighter one- and two-photon confocal images. Before counting, the background intensity is estimated by a smoothing procedure from the data and then included in our refined model. The result is a segmentation map of the image area shown in Fig.~\ref{fig:origami-measurement}b where an estimated number of molecules as well as a confidence interval on the number of molecules is assigned to each segment. As expected the estimated number of molecules is always within the calculated confidence bounds. The width of the confidence bounds ($K$) is about as large as the estimated number of molecules.
	
	\section{Discussion and outlook}\label{sec:discussion}
	
	\subsection{Summary}
	
	In this work, a comprehensive physical and statistical modeling of the coincidence photon statistics encountered in antibunching microscopy has been conducted. The coincidence signals recorded from parallel detectors in a high-resolution fluorescence microscope intricately depend on the number of fluorescent molecules at a specific location as well as their brightness. Under the mild assumption that molecules locally exhibit the same brightness, and under neglecting all background contributions, a rigorous statistical model of the problem of estimating the number of molecules within small regions of the sample has been developed. Based on this an estimation procedure of the number of molecules was presented and was shown to converge to the true molecule counts for sufficiently long measurement times. Furthermore, confidence intervals for the local number of molecules have been constructed. The estimation procedure requires to choose regions of the sample in which molecule numbers are to be estimated. A hybrid-segmentation has been introduced, which combines a natural segmentation approach with guarantees that all segments contain molecules with high probability. In a numerical study, the convergence of the estimator has been affirmed. With increasing number $t$ of excitation pulses the obtained segment sizes as well as the size of the confidence intervals have been shown to shrink.
	
	\subsection{Future work}
	
	In previous works the number of molecules (as well as their brightness) was estimated as a continuous density from a global fit of the data with a coincidence statistics model up to second order (\citet{t15}). The model presented here includes arbitrarily high photon coincidence orders. However, in this work only the lowest two orders of $\hat s_k(\bx_\bi)$ were used in the estimation of the local number of molecules. The optimal choice of the used number of orders depends on the accuracy of the estimation of each $\hat s_k$ and may be subject to further studies. An important issue here is the quality of the available data (i.e. the possible number of excitation pulses $t$), as otherwise the higher order photon coincidences will not be able to provide additional information.
	
	Segmentation-based estimation of local molecular counts was essential for the generation of confidence information. The importance of choosing an appropriate sample segmentation which provides statistically sound guarantees (see Theorem~\ref{thm:seg}) for generating useful confidence information is a key finding of this work. A very large number of localized segments would result in overly large confidence intervals (due to multiplicity), while a small number of extended segments would not result in meaningful local information. A balanced approach as suggested in this work results in useful molecule counting information. A multitude of different segmentations are imaginable but the hybrid segmentation used in this work appears to be working reasonably well. Further work might deliver more insights into how segmentations of the sample can maximize the usefulness of the counting information in these cases.
	
	The enlarged segments used for estimation are in principle required to contain all the collected fluorescence from all molecules within the segment and none from molecules outside the segment (segments should contain a molecule-free border with a width of at least the FWHM of the PSF). Clearly, this is not always the case in practice, and is often difficult to achieve in densely labelled samples unless one restricts oneself to rather large segments, which would not result in useful local information. This will in general result in an uncontrollable bias in the molecule number estimation. In the case of multiple closely spaced segments, lost photon contributions from molecules within a segment can be partially compensated by contributions from molecules in adjacent segments. In our examples (see Figs.~\ref{fig:workflow} and \ref{fig:two-clusters}) the estimated confidence intervals were not compromised. A further careful analysis of segment border effects might result in additional contributions to the confidence interval sizes.
	
	
	The analysis performed in this study also assumed that molecules remain intact during the whole measurement. Indeed, in every microscopy experiment, fluorophores will eventually lose their ability to fluoresce (photobleach). To minimize estimation biases, we restricted the duration of the experiment to the characteristic time that will leave a strong majority of fluorophores intact. Should, nevertheless, some molecules photobleach during an experiment, we expect the number estimation to effectively approximate the mean number of unbleached molecules during the measurement. Photobleaching could be included in a more refined statistical model.
	
	
	As discussed in the simulations, a too large number of molecules (or more precisely a too large product of local number of molecules and their brightness) in comparison to the number $\md$ of detectors used yields a systematic under-estimation of the number of molecules. This is due to the fact the bias introduced by the model from Theorem~\ref{Le:trans} is no longer negligible. Our current experimental setup shown in Figure~\ref{fig:experimental_setup} using four detectors is able to count approximately 50 molecules at a single diffraction limited spot accurately. Increasing the number of detectors modestly would increase this limit significantly as shown in Figure~\ref{fig:single-cluster-vs-md}. Further advances in detector technology currently being invented (being able to count single photons without significant dead times and therefore not limiting the observable coincidence photon order) are expected to diminish if not solve the problem completely while at the same time simplifying the microscope setup further.
	
	All given examples in this work have been executed in two dimensions only, which corresponds to an often used imaging mode in scanning microscopy. However, the extension to 3D is straightforward and does not require any particular effort.
	
	
	\begin{appendix}
		\section{Proofs}\label{app:proofs}
		
	\subsection{Proof of Theorem 2.1}

\begin{lem}\label{le:Aandg}
	In the setting of Theorem 
	2.1, there exists a linear map $A$ and a non-linear transformation $g$ such that $\trans=A\circ g$.
\end{lem}

\begin{proof}[Proof of Lemma \ref{le:Aandg}]
	Due to superposition and independence of the markers, for each position $\bx$ the random variable that returns the number of photons emitted after one excitation pulse follows a discrete probability distribution of a sum of independent Bernoulli trials that are not necessarily identically distributed. This distribution is called \textit{Poisson binomial distribution} \citep{Tang2019ThePB}. The probability that exactly $k$ photons are emitted  at scan position $\bx$ in one experiment is denoted by $Q_k'(\bx)$, which is given by
	\begin{align*}
		Q_0' \left(\bx\right) &= \prod\limits_{j=1}^N \left(1-\brightx_j\left(\bx\right)\right), \\ 
		\quad Q_k' \left(\bx\right) &= \sum\limits_{1 \leq i_1 < ... < i_k \leq N} \prod\limits_{j=1}^k \brightx_{i_j} \left(\bx\right) \prod\limits_{j \neq i_1,...,i_k} \left(1- \brightx_j \left(\bx\right) \right),
	\end{align*}
	for $k\in\{1,\ldots,N\}$.
	Note, that for $k>1$  $Q_k'\left(\bx\right)$ requires the different photons to originate from different markers. In practice, we do not have direct empirical access to the probabilities $Q_k'$. To see this, suppose for instance that $\md = 4$ as in Figure 
	1 in the main document, and that $2$ photons were emitted. If the beam splitters in our experimental setup distribute the incoming photons equally likely in each direction, then after the first beam splitter with probability $50 \%$, the photons have been separated and will arrive at different detectors. On the other hand, in $50 \%$ of the cases, both photons have been sent in the same direction and will arrive together at the next beam splitting. Continuing this argument, it follows that in $75\%$ of all cases, the two photons are sent towards different detectors, whereas in $25\%$ of all cases, the two photons are sent towards a single detector, which cannot differentiate the number of incident photons. Therefore, with
	\begin{align*}
		D_i(\bx)&:=\mathbb{P}\left(i \;\text{active detectors at } \bx\; \text{in one experiment}\right),
	\end{align*}
	the probability $Q_2' \left(\bx\right)$ distributes with weight $1/4$ to the probability $D_1 \left(\bx\right)$ to observe a single active detector, and with weight $3/4$ to the probability $D_2 \left(\bx\right)$ to observe exactly two active detectors. 
	The probabilities $D_i$ can be computed in terms of the $Q_i'$s as
	\begin{align*}
		D_i(\bx)
		&=\sum_{j=i}^N\mathbb{P}\left(i\;\text{active detectors}\,|\,j\;\text{photons}\right)\times\mathbb{P}\left(j\;\text{photons}\right)\\
		&=\sum_{j=i}^N\mathbb{P}\left(i\;\text{active detectors}\,|\,j\;\text{photons}\right)Q_j'(\bx).
	\end{align*}
	Given that $j$ photons are emitted, there are in total $\md^j$ ways to distribute those onto $\md$ detectors. The total number of outcomes in which $0 \leq i \leq \md$ detectors are activated is given by the product of the number of possibilities to distribute $j$ photons onto one specific selection of $i$ detectors and the number of different choices of $i$ out of $\md$ detectors. The first factor is given by $S(j,i)\times i!$, where $S(j,i)$ denote the Stirling numbers of the second kind\footnote{The Stirling number $S(j,i)$ is   the number of ways to partition a set of $j$ objects into $i$ non-empty subsets.}. The second factor is given by the binomial coefficient $\binom{\md}{i}$. Since the emitted photons are distributed with equal probability onto $\md$ detectors, we obtain
	\begin{align*}
		\mathbb{P}\left(i\;\text{active det.}\,|\,j\;\text{photons}\right)=\frac{S(j,i)\cdot i!\cdot\binom{\md}{i}}{\md^j}=\frac{S(j,i)(\md-1)!}{(\md-i)!\md^{j-1}}=:w_{i,j}.
	\end{align*} 
	This yields 
	\begin{align*}
		D_i(\bx)=\sum_{j=i}^{N}Q_{j}'(\bx)w_{i,j}, 
	\end{align*}
	which yields that $D_0(\bx)=Q_0'(\bx)$, meaning that whenever at least one photon is emitted, at least one detector will be active.
	Moreover, the contribution of summands for large values of $j$, $j>\md$ say, are negligible. Hence, we write
	\begin{align*}
		D_i(\bx)=\sum_{j=i}^{\md}Q_{j}'(\bx)w_{i,j}+\sum_{j=\md+1}^{N}Q_{j}'(\bx)w_{i,j}.
	\end{align*}
	Let $\mathbf{Q}(\bx)=(Q_1'(\bx),\ldots,Q_{\md}'(\bx))^T$ denote the vector of the first $m_d$ probabilities $Q_k'$ and let $\mathbf{D}(\bx)=(D_1(\bx),\ldots,D_{\md}(\bx))^T$. Then, for $i>0$, we obtain the following connection between $\mathbf{D}$ and $\mathbf{Q}$ 
	\begin{align}\label{eq:DQ}
		\mathbf{D}(\bx)=A\mathbf{Q}(\bx)+ \sum_{j=\md+1}^{N}Q_{j}'(\bx)\mathbf{w}_{j},
				%
				%
				%
	\end{align}
	with
	\begin{align*}
		A=\left(w_{p,q}I\{p\leq q\}\right)_{p,q=1}^{\md}
	\end{align*}
	and $\mathbf{w}_{j}=(w_{1,j},\ldots,w_{N,j})$.
		The triangular matrix $A$ satisfies $\det(A)=\prod_{l=1}^{\md}w_{l,l}>0$ and therefore, the matrix  $A$ is invertible.
		
		Next, we derive  representations of the probabilities $Q_{k}'$ in terms of 
		\begin{align*}
			s_k\left(\bx\right) &:= \sum\limits_{i=1}^N \brightx_i\left(\bx\right)^k, \qquad k \in \mathbb N, \\
			S_k \left(\bx\right)&:= \sum\limits_{i_1, ..., i_k \in \left\{1,...,N\right\} \atop i_1 \neq ... \neq i_k} \brightx_{i_1} \left(\bx\right) \cdot ... \cdot \brightx_{i_k}\left(\bx\right), \qquad k \in \mathbb N_0,
		\end{align*}
		which give rise  to
		\begin{equation}\label{eq:Q_k'}
			Q_k'(\bx)  =  \frac{1}{k!} \sum\limits_{j=0}^{N-k} \frac{\left(-1\right)^{j}}{j!} S_{j+k}(\bx), \qquad 0 \leq k \leq \md.
		\end{equation}
		Here, the factors $1/k!$ and $1/j!$ arise from the number of possible orderings when replacing $\left\{1 \leq i_1 < ... < i_k\leq N\right\}$ with $\left\{i_1,...,i_k~|~ i_1 \neq ... \neq i_l\right\}$, respectively.
		\,Note that the iterated sums $S_k \left(\bx\right)$ can be computed recursively, requiring knowledge of the $s_k(\bx)$ only, by the formula
		\begin{align}
			S_0 &= 1, \notag\\
			S_k&= \sum\limits_{j=1}^k \left(-1\right)^{j+1} \frac{\left(k-1\right)!}{\left(k-j\right)!} s_jS_{k-j},\label{eq:formula_S_k} 
		\end{align}
		Define the nonlinear transformation $g : \mathbb R^{\md} \to \mathbb R^{\md}$ by
		\[
		g \left(s_1(\bx),...,s_{\ms}(\bx)\right) =  \left(\frac{1}{k!}\sum\limits_{j=0}^{\md-k} \frac{\left(-1\right)^{j}}{j!} S_{j+k}(\bx) \right)_{1 \leq k \leq \md}=:\left(\widetilde{Q}_k(\bx) \right)_{1 \leq k \leq \md},
		\]
		with $S_j(\bx)$ as in \eqref{eq:formula_S_k}. 
		Furthermore, we find
		\begin{multline*}
			A\left(g(\bs(\bx))\right)=A\mathbf{Q}(\bx)-A \left(\frac{1}{k!}\sum\limits_{j=\md-k+1}^{N-k} \frac{\left(-1\right)^{j}}{j!} S_{j+k}(\bx)\right)_{1\leq k\leq\md}
			\\
			=\bD(\bx)-A \left(\frac{1}{k!}\sum\limits_{j=\md-k+1}^{N-k} \frac{\left(-1\right)^{j}}{j!} S_{j+k}(\bx)\right)_{1\leq k\leq\md}-\sum_{j=\md+1}^{N}Q_{j}'(\bx)\mathbf{w}_{j}.
		\end{multline*}
		Since
		\begin{align*}
			\left\| \sum_{j=\md+1}^{N}Q_{j}'(\bx)\mathbf{w}_{j}\right\|_{2}&\leq\sum_{j=\md+1}^{N}\gamma^{j}\|\mathbf{w}_j\|_2\leq\sqrt{\md}\sum_{j=\md+1}^{N}\gamma^{j},\\
			&=\sqrt{\md}\frac{1-\gamma^{N-\md}}{1-\gamma}\cdot\gamma^{\md+1},
		\end{align*}
		$\|A\|_2\leq1$
		and, for any $k$,
		\begin{align*}
			\frac{1}{k!}\sum\limits_{j=\md-k+1}^{N-k} \frac{\left(-1\right)^{j}}{j!} S_{j+k}(\bx)  \leq\frac{e-1}{k!}\gamma^{\md+1}
		\end{align*}
		
		the claim of  Lemma \ref{le:Aandg} now follows.
		
	\end{proof}
	To conclude the proof of Theorem 
	2.1, it remains to show that the map $g$ is invertible.
	Given $\widetilde{Q}_k, k=1,\ldots,\md$, the quantities $S_1,\ldots,S_{\md}$ can be recovered as follows. Starting with $k=\md,$ we obtain $$\widetilde{Q}_{\md}(\bx)\cdot\md!=S_{\md}(\bx).$$ Next, for $k=\md-1$, we obtain $$\widetilde{Q}_{\md-1}(\bx)=\frac{1}{(\md-1)!}\cdot(S_{\md}(\bx)+S_{\md-1}(\bx)),$$
	yielding
	$$(\md-1)!\cdot\widetilde{Q}_{\md-1}(\bx)-S_{\md}(\bx)=S_{\md-1}(\bx)).$$
	Successively, $S_{\md-2},\ldots,S_1$ can be recovered in the same fashion.
	Given $S_1,\ldots,S_{\md}$, we can compute $s_1,\ldots,s_{\md}$ successively as well. Starting with $k=1$, we find $S_1=s_1\cdot S_0=s_1.$ Next, for $k=2$, we find $S_2=s_1S_1-s_2S_0$, which gives $s_2=s_1^2-S_2.$ One by one, $s_3,\ldots,s_{\md}$ can be recovered.%
	\hfill$\Box$

	\subsection{Proof of Theorem 
		3.1}
	
	\begin{enumerate}
		\item It is clear that with $\widetilde{\mathcal{B}}$ satisfying 
		\begin{equation}\label{eq:FWER}
			\inf_{C\in\mathcal{C}_N}\mathbb P \left[f_{|_{B_{\bx,\bh}}} \not\equiv 0 \text{ for all } B_{\bx,\bh} \in \widetilde{\mathcal{B}}\,|\,\text{configuration }C_N\right] \geq 1-\alpha,
		\end{equation}
		also any set $\mathcal{B}\subset \widetilde{\mathcal{B}}$ satisfies \eqref{eq:FWER}. The construction of $\widehat{\mathcal{ROI}}$ ensures that it only contains supersets of sets in $\mathcal{B}$, such that $\widehat{\mathcal{ROI}}$ inherits \eqref{eq:FWER}.
		\item This is clear by construction, as intersecting sets are always merged. 
	\end{enumerate}
	
	\subsection{Proof of Theorem 
		4.2}
	We will give the proof for the case of $|R|=1$. While it immediately extends to the case $|R|>1$, it provides a substantial ease of notation.
	In the proof of Theorem
	2.1, the matrix $A$ and the function $g$ have been defined for the case $N>\md$ as this is the relevant case in practice. However, for our asymptotic considerations, we need to extend these quantities to the case of $N\leq\md$. Since these definitions are somewhat artificial and are only relevant within this proof, we will denote them as $\At$ and $\gt$, respectively. 
	First, we define the matrix $\At$. To this end, let $a_{p,q}=0$, if $q<p$ or if $p<q$ and $q>N$ and  let $a_{p,q}=w_{p,q}$ else. Then $\At=(a_{p,q})_{p,q=1,\ldots,\md}$. With this definition, the matrix $\At$ is an upper triangular matrix with positive diagonal elements and block structure
	\begin{align}\label{eq:block}
		\At=\left(\begin{array}{cc}
			A_1 & A_2  \\
			A_3 & A_4 
		\end{array}
		\right),
	\end{align}
	where $A_1\in\R^{N\times N}$ and $A_4\in\R^{(\md-N)\times(\md-N)}$ are upper triangular matrices with positive diagonal elements,  $A_2=\mathbf{0}\in\R^{N\times(\md-N)}$ and $A_4=\mathbf{0}\in\R^{(\md-N)\times N}.$ Therefore, $\At^{-1}$ exists and has the same block structure. Now, in order to define the function $\gt,$ let
	\begin{align*}
		b_{i,j}=\begin{cases}
			\frac{1}{i!}\frac{(-1)^{j-i}}{(j-i)!} & i=j\;\text{or}\;j>i\;\text{and}\;j\leq N\\
			0 & \text{else}
		\end{cases}.
	\end{align*}
	Then, the matrix $B:=(b_{i,j})_{i,j=1,\ldots,\md}$ has the same structure as the matrix $\At$ in \eqref{eq:block}. In particular, $B$ is invertible with $B^{-1}=(\beta_{i,j})_{i,j=1,\ldots,\md}$. Define
	\begin{align*}
		\gt(s_1(\bx),\ldots,s_{\md}(\bx))=B\cdot(S_1(\bx),\ldots,S_{\md}(\bx))^T,
	\end{align*}
	where the relation between the $s_k$'s and the $S_k$'s is given in \eqref{eq:formula_S_k}. 
	Notice that $\At=\A$ and $\gt=\g$ if $N>\md$. 
	If $\md>N$
	\begin{align*}
		\gt(\mathbf{s}(\bx))=\left(\begin{array}{c}
			Q_1'(\bx)\\
			\vdots\\
			Q_N'(\bx)\\
			0\\
			\vdots\\
			0
		\end{array}\right)=:\mathbf{Q}^*(\bx)
	\end{align*}
	$\gt$ is invertible, which can be shown analogously to the case $N>\md$.
	If $|R|=1$, the map $\transclt$ becomes
	\begin{align*}
		\transclt(\by):\begin{cases}
			\R^{\md}\to\R\\
			\by\mapsto\frac{H_2}{H_1^2}\frac{(\Pi^{\md}_1\gt^{-1}\A_t^{-1}\by)^2}{\Pi^{\md}_2\gt^{-1}\A_t^{-1}\by}
		\end{cases}
	\end{align*}
	with 
	\begin{align*}
		\frac{H_1^2}{H_2}\nabla\Psi(\by)=&2\frac{\Pi^{\md}_1\gt^{-1}\A_t^{-1}\by}{\Pi^{\md}_2\gt^{-1}\A_t^{-1}\by}\nabla \big(\Pi^{\md}_1\gt^{-1}\A_t^{-1}\big)(\by)
		\\&-\frac{(\Pi^{\md}_1\gt^{-1}\A_t^{-1}\by)^2}{(\Pi^{\md}_2\gt^{-1}\A_t^{-1})^2\by}\nabla \big(\Pi^{\md}_2\gt^{-1}\A_t^{-1}\big)(\by),
	\end{align*}
	where
	\begin{align*}
		\nabla \big(\Pi^{\md}_1\gt^{-1}\A_t^{-1}\big)(\by) =\left(\big(J_{\gt^{-1}}(\At^{-1}\by)\big)_{1}A_t^{-1}\right)^T 
	\end{align*}
	and
	\begin{align*}
		\nabla \big(\Pi^{\md}_2\gt^{-1}\A_t^{-1}\big)(\by) =\left(\big(J_{\gt^{-1}}(\At^{-1}\by)\big)_{2}A_t^{-1}\right)^T . 
	\end{align*}
	By $\big(J_{\gt^{-1}}(\At^{-1}\by)\big)_{k}$ we denote the $k$'th row of the matrix $J_{\gt^{-1}}(\At^{-1}\by)$.
	\noindent Since there is at least one marker in the region $R$ by assumption, we have for one $\bx\in R$
	\begin{align*}
		S_1(\bx)=s_1(\bx)=\sum_{j=1}^{\md}\beta_{1,j}Q_j(\bx)=\left(\gt^{-1}\right)_{1}(\mathbf{Q}^*(\bx))>0,
	\end{align*}
	such that $s_2(\bx)$ is well defined:
	\begin{multline*}
		s_2(\bx)=\frac{S_1(\bx)-S_2(\bx)}{S_1(\bx)}\\=\frac{\sum_{j=1}^{\md}\beta_{1,j}Q_j(\bx)-\sum_{j=1}^{\md}\beta_{2,j}Q_j(\bx)}{\sum_{j=1}^{\md}\beta_{1,j}Q_j(\bx)}=\left(\gt^{-1}\right)_2(\mathbf{Q}^*(\bx)).
	\end{multline*}
	This yields
	\begin{align*}
		\left(J_{\gt^{-1}}(\mathbf{Q}^*(\bx)\right)_1=(\beta_{1,1},\ldots,\beta_{1,N},0,\ldots,0)
	\end{align*}
	and
	\begin{align*}
		\left(J_{\gt^{-1}}(\mathbf{Q}^*(\bx)\right)_2 =\left(\widetilde{\beta}_1,\ldots,\widetilde{\beta}_N,0,\ldots,0\right),
	\end{align*}
	where
	\begin{align*}
		\widetilde{\beta}_k=\frac{\beta_{1,k}-\beta_{2,k}}{\sum_{j=1}^{\md}\beta_{1,j}Q_j}
		-\beta_{1,k}\frac{\sum_{j=1}^{\md}\beta_{1,j}Q_j-\sum_{j=1}^{\md}\beta_{2,j}Q_j}{\left(\sum_{j=1}^{\md}\beta_{1,j}Q_j\right)^2}.
	\end{align*}
	In particular, only the first $N$ entries are non-zero. Therefore, multiplication with the matrix
	\begin{align*}
		\At^{-1}=\left(\begin{array}{cc}
			\widetilde A_1 & \widetilde A_2  \\
			\widetilde A_3 & \widetilde A_4 
		\end{array}
		\right)
	\end{align*}
	with $\widetilde A_2=\mathbf{0}\in\R^{(\md-N)\times N}$ yields again an element with zero-entries after the $N$-th component.
	Since the model bias vanishes if $N\leq\md$, 
	we find
	\begin{align}\label{eq:Ft}
		F_t(s):= &\mathbb{P}\left(\sqrt{t}\Big(\widehat N_R-N_R\Big)\leq s\right)=
		\mathbb{P}\left(\sqrt{t}(\transcltDhat-\transcltD)\leq s\right)\\
		&=\mathbb{P}\left(\sqrt{t}\nabla\transcltD^T(\widehat{\mathcal{E}}-\mathcal{E})+
		\frac{\sqrt{t}}{2}(\widehat{\mathcal{E}}-\mathcal{E})^T\mathrm{Hess}_{\transclt}(\widetilde{\mathcal{E}})(\widehat{\mathcal{E}}-\mathcal{E})\leq s\right),\notag
	\end{align}
	for an intermediate point $\widetilde{\mathcal{E}}$. Since $\|\widehat{\mathcal{E}}-\mathcal{E}\|_2\leq\|\widehat{\mathcal{E}}-\mathcal{E}\|_1$, where $\|\cdot\|_1$ and $\|\cdot\|_2$ denote the $l^1-$ and $l^2-$norm on $\mathbb{R}^{|R|\md}$, respectively, we find that
	\begin{align*}
		\mathbb{P}\left(\|\widehat{\mathcal{E}}-\mathcal{E}\|_2^2>\varepsilon\right)\leq  \mathbb{P}\left(\|\widehat{\mathcal{E}}-\mathcal{E}\|_1>\sqrt{\varepsilon}\right)\leq3\exp(-t\varepsilon/25),
	\end{align*}
	where the last inequality follows from Lemma 3 in \cite{Devroye1983}, using $\md\leq5\log(t)$. Setting $\varepsilon=\log(t)/t\cdot25$ gives
	\begin{align*}
		\mathbb{P}\left(\|\widehat{\mathcal{E}}-\mathcal{E}\|_2^2>\frac{25\log(t)}{t}\right)\leq \frac{3}{t}. 
	\end{align*}
	By assumption of the theorem, there exists a positive constant $C>0$ such that
	\begin{align*}
		|(\widehat{\mathcal{E}}-\mathcal{E})^T\mathrm{Hess}_{\transclt}(\widetilde{\mathcal{E}})(\widehat{\mathcal{E}}-\mathcal{E})|\leq C\|\widehat{\mathcal{E}}-\mathcal{E}\|_2^2.
	\end{align*}
	Hence, with $F_t$ defined in \eqref{eq:Ft}, we find
	\begin{align*}
		F_t^-(s)\leq F_t(s)\leq
		F_t^+(s),
	\end{align*}
	where
	\begin{align*}
		F_t^+(s)&:=\mathbb{P}\left(\sqrt{t}\nabla\transcltD^T(\widehat{\mathcal{E}}-\mathcal{E})
		\leq s+\frac{C\sqrt{t}}{2}\|\widehat{\mathcal{E}}-\mathcal{E}\|_2^2\right)\\
		&\leq \mathbb{P}\left(\sqrt{t}\nabla\transcltD^T(\widehat{\mathcal{E}}-\mathcal{E})
		\leq s+25C\frac{\log(t)}{2\sqrt{t}}\right)+\frac{3}{t},
	\end{align*}
	as well as
	\begin{align*}
		F_t^-(s)&:=\mathbb{P}\left(\sqrt{t}\nabla\transcltD^T(\widehat{\mathcal{E}}-\mathcal{E})
		\leq s-\frac{C\sqrt{t}}{2}\|\widehat{\mathcal{E}}-\mathcal{E}\|_2^2\right)\\
		&\geq \mathbb{P}\left(\sqrt{t}\nabla\transcltD^T(\widehat{\mathcal{E}}-\mathcal{E})
		\leq s-25C\frac{\log(t)}{2\sqrt{t}}\right)-\frac{3}{t},
	\end{align*}
	For given $v\in\mathbb{R}^{\md}$ and $z\in\mathbb{R}$,  the set
	\begin{align*}
		A_{v,z}=\{w\in \mathbb{R}^{|R|\md}\,|\,v^Tw\leq z\}
	\end{align*}
	defines a closed half space in $\mathbb{R}^{|R|\md}$ and as such, it is a 1-generated, closed convex set and we have
	\begin{align*}
		\mathbb{P}\left(\sqrt{t}\nabla\transcltD^T(\widehat{\mathcal{E}}-\mathcal{E})
		\leq s-\tau_t\right)=\mathbb{P}\left(\sqrt{t}(\widehat{\mathcal{E}}-\mathcal{E})
		\in A_{\transcltD,s-\tau_t}\right),
	\end{align*}
	where $\tau_t=25C\frac{\log(t)}{2\sqrt{t}}$. Note that we can write
	\begin{align*}
		(\widehat{\mathcal{E}}-\mathcal{E})=\frac{1}{t}\sum_{k=1}^{t}
		\left[
		\left(\begin{array}{c}
			I\{M_k(\bx_{\bi_1})=0\}\\\vdots\\I\{M_k(\bx_{\bi_1})=\md\}
		\end{array}
		\right)-\left(\begin{array}{c}
			D_0(\bx_{\bi_1})\\\vdots\\ D_{\md}(\bx_{\bi_1})
		\end{array}
		\right)
		\right]
		=\frac{1}{t}\sum_{k=1}^{t}\mathcal{M}_k,
	\end{align*} 
	where $\mathcal{M}_k$ is defined in an obvious manner.
	It follows from Proposition 3 in \cite{CCK2017} that if the conditions M.1', M.2' and E.2' hold, where
	
	\begin{align*}
		&(M.1')\quad \frac{1}{t}\sum_{j=1}^t\mathbb{E}\left(\frac{\nabla\transcltD^T}{\|\nabla\transcltD\|}\cdot \mathcal{M}_j\right)^2\geq b\quad\text{for some positive constant $b$,}\\
		&(M.2')\quad \frac{1}{t}\sum_{j=1}^t\mathbb{E}\left|\frac{\nabla\transcltD^T}{\|\nabla\transcltD\|}\cdot \mathcal{M}_j\right|^{2+k}\leq B_t^k,\quad k=1,2,\\
		&(E.2')\quad \mathbb{E}\left[\exp\left(\left|\frac{\nabla\transcltD^T}{\|\nabla\transcltD\|}\cdot \mathcal{M}_j\right|/B_t\right)\right]\leq2\quad\forall j=1,\ldots, t,
	\end{align*}
	we obtain 
	\begin{multline*}
		\sup_{s\pm\tau_t\in\mathbb{R}}\left|\mathbb{P}\left(\sqrt{t}(\widehat{\mathcal{E}}-\mathcal{E})
		\in A_{\transcltD,s\pm\tau_t}\right)-\mathbb{P}\left(\widetilde{Z}
		\in A_{\transcltD,s\pm\tau_t}\right)\right|\\
		=\sup_{s\in\mathbb{R}}\left|\mathbb{P}\left(\sqrt{t}(\widehat{\mathcal{E}}-\mathcal{E})
		\in A_{\transcltD,s\pm\tau_t}\right)-\mathbb{P}\left(\widetilde{Z}
		\in A_{\transcltD,s\pm\tau_t}\right)\right|\\\leq C\left(\frac{B_t^2\log(\md|R| t)}{t}\right)^{\frac{1}{6}},
	\end{multline*}
	where $\widetilde Z\sim\mathcal{N}(0,\Sigma_R)$. We now show that conditions $(M.1'), (M.2')$ and $(E.2')$ are satisfied with $B_t=\md^{\frac{3}{2}}$.
	\smallskip
	
	\noindent\textbf{Verifying condition (M.1'):}\\
	Since the $\mathcal{M}_k$ are i.i.d., we find 
	\begin{align*}
		\frac{1}{t}\sum_{j=1}^t\mathbb{E}\left(\frac{\nabla\transcltD^T}{\|\nabla\transcltD\|}\cdot \mathcal{M}_j\right)^2=\mathbb{E}\left(\frac{\nabla\transcltD^T}{\|\nabla\transcltD\|}\cdot \mathcal{M}_1\right)^2.
	\end{align*}
	Both $\nabla\Psi$ and $\mathcal{M}_1$ only have at most $N$ non-zero entries (and $N$ does not depend on $t$).
	Let $v_t\in \mathbb{R}^{\md|R|}$, $\|v_t\|=1$, be a deterministic vector. Then, if and only if the realizations of $\mathcal{M}_1$ are collinear on an event $\Omega_{0,t}$ with $\mathbb{P}(\Omega_{0,t})\to1$ as $t\to\infty$
	\begin{align*}
		\lim_{t\to \infty}\mathbb{E}\Big[\big(v_t^T\mathcal{M}_1\big)^2\Big]= 0 .
	\end{align*}
	
	By assumption, there is at least one marker in the region $R$ with individual brightness smaller than 0.5. Therefore,
	there exists a positive constant $\tilde b>0$ such that  $(D_0(\bx),D_1(\bx))\neq(0.5,0.5)$ and $D_0(\bx)\wedge D_1(\bx)>\tilde b$. The first two components of the random vector $\mathcal{M}_1$ can either be 
	\begin{align*}
		\mathbf{u}_1:=\binom{1-D_0(\bx)}{-D_1(\bx)}\quad\text{or}\quad \mathbf{u}_2:=\binom{D_0(\bx)}{1-D_1(\bx)}.
	\end{align*} 
	The probabilities for both events are bounded away from $0$ as $t\to\infty$. Vectors $\mathcal{M}_1$ with first two components equal to $\mathbf{u}_1$ cannot be collinear to vectors $\mathcal{M}_1$ with first two components equal to $\mathbf{u}_2$. To see this,
	let $\beta\in\mathbb{R}.$ Then
	\begin{align*}
		\binom{1-D_0(\bx)}{-D_1(\bx)}=\beta \binom{D_0(\bx)}{1-D_1(\bx)}
	\end{align*}     
	iff 
	\begin{align*}
		\beta=\frac{1-D_0(\bx)}{D_0(\bx)}  \quad\text{and}\quad  \beta=\frac{D_1(\bx)}{1-D_1(\bx)}.
	\end{align*}
	The latter condition can only be satisfied if 
	$(D_0(\bx_{\bi_j}),D_1(\bx_{\bi_j}))=(0.5,0.5)$. Therefore, at least one of the vectors containing either $\mathbf{u}_1$ or $\mathbf{u}_2$ is not perpendicular to $v_t$, such that condition (E.2') is satisfied.
	%
	\textbf{Verifying condition (M.2'):}\\
	We have that
	\begin{multline*}
		\frac{1}{t}\sum_{j=1}^t\mathbb{E}\left|\frac{\nabla\transcltD^T}{\|\nabla\transcltD\|}\cdot \mathcal{M}_j\right|^{2+k}=\mathbb{E}\left|\frac{\nabla\transcltD^T}{\|\nabla\transcltD\|}\cdot \mathcal{M}_1\right|^{2+k}\\\leq\frac{\|\nabla\transcltD\|_1^{2+k}}{\|\nabla\transcltD\|_2^{2+k}}\leq (\md|R|)^{\frac{2+k}{2}}.
	\end{multline*}
	with $B_t=(\md|R|)^{\frac{3}{2}}$.\\
	\textbf{Verifying condition (E.1'):}\\
	We need to show that
	$$\mathbb{E}\left[\exp\left(\left|\frac{\nabla\transcltD^T}{\|\nabla\transcltD\|}\cdot \mathcal{M}_1\right|/B_t\right)\right]\leq2.$$  Since $\left|\frac{\nabla\transcltD^T}{\|\nabla\transcltD\|}\cdot \mathcal{M}_1\right|\leq B_t^{1/3}$, condition (E.2') trivially holds for sufficiently large $t$.
	Therefore, all three conditions are met and we obtain the following statement
	\begin{multline*}
		\sup_{s\in\mathbb{R}}\left|\mathbb{P}\left(\sqrt{t}(\widehat{\mathcal{E}}-\mathcal{E})
		\in A_{\transcltD,s+\tau_t}\right)-\mathbb{P}\left(\widetilde{Z}
		\in A_{\transcltD,s+\tau_t}\right)\right|\\\leq C\left(\frac{(\md|R|)^{3}\log(\md|R|t)}{t}\right)^{\frac{1}{6}}.
	\end{multline*} 
	Finally, we show that $\mathbb{P}\left(\widetilde{Z}
	\in A_{\transcltD,s+\tau_t}\right)$ and $\mathbb{P}\left(\widetilde{Z}
	\in A_{\transcltD,s}\right)$ are close, uniformly in $s$.
	Since $\widetilde Z\sim\mathcal{N}(0,\nabla\Psi(\mathcal{E})^T \Sigma_R\nabla\Psi(\mathcal{E}))$, it follows that
	
	\begin{multline*}
		\sup_{s\in\mathbb{R}}\left|\mathbb{P}\left(\widetilde{Z}
		\in A_{\transcltD,s+\tau_t}\right)-\mathbb{P}\left(\widetilde{Z}
		\in A_{\transcltD,s}\right)\right|\\\leq\sup_{s\in\mathbb{R}}\left|\Phi\left(\frac{s}{\sigma_R}\right)-\Phi\left(\frac{s+\tau_t}{\sigma_R}\right)\right|\leq\frac{1}{\sqrt{2\pi}}\frac{\tau_t}{\sigma_R}.
	\end{multline*} 
	Combining all previous steps yields
	\begin{multline*}
		\sup_{s\in\mathbb{R}} \left|\mathbb{P}\left(\sqrt{t}\Big(\widehat N_R-N_R\Big)\leq s\right)-\mathbb{P}\left(\widetilde{Z}
		\in A_{\transcltD,s}\right)\right|\\\leq \widetilde C\left(\frac{1}{t}+\frac{\log(t)}{\sqrt{t}}+\left(\frac{(\md|R|)^{3}\log(\md|R| t)}{t}\right)^{\frac{1}{6}}\right).
	\end{multline*}
	With $\md\leq5\log(t)$ the claim of the theorem now follows.
	
	\hfill$\Box$
		
	\end{appendix}

	\section*{Acknowledgements}
	The authors want to thank Stefan W. Hell for helpful discussions. All authors gratefully acknowledge financial support by the DFG through CRC 755, project A07. F. W. is furthermore supported by the DFG via grant WE 6204/4-1. The authors also want to thank two anonymous referees for valuable comments which helped to improve the presentation of the manuscript.
	
	\bibliographystyle{apalike} %
	\bibliography{cfmbib}     
	
\end{document}